# A global Canopy Water Content product from AVHRR/Metop

Francisco Javier García-Haro[a*], Manuel Campos-Taberner[a], Álvaro Moreno[b,e], Håkan Torbern Tagesson[c], Fernando Camacho[d], Beatriz Martínez[a], Sergio Sánchez[a], María Piles[b], Gustau Camps-Valls[b], Marta Yebra[f,g] María Amparo Gilabert[a]

[a]Environmental Remote Sensing group (UV-ERS),
Universitat de València, Dr. Moliner, 46100 Burjassot, València (Spain).

[b]Image Processing Laboratory (IPL), Universitat de València, C/ Catedrático José Beltrán, 2, 46980 Paterna, València (Spain).

[c] Department of Geosciences and Natural Resource Management (IGN), University of Copenhagen, Copenhagen, Denmark.

[d]Earth Observation Laboratory (EOLAB), Parc Científic de la Universitat de València, Catedrático A. Escardino, 46980 Paterna, València (Spain).

[e]Numerical Terradynamic Simulation Group (NTSG), College of Forestry and Conservation, University of Montana, Missoula, USA

[f]Fenner School of Environment and Society, The Australian National University, ACT, Canberra, Australia
[g]Bushfire & Natural Hazards Cooperative Research Centre, Melbourne, Australia

[*]E-mail: j.garcia.haro@uv.es; Tel. +34 963543111

## Abstract

Spatially and temporally explicit canopy water content (CWC) data are important for monitoring vegetation status, and constitute essential information for studying ecosystem-climate interactions. Despite many efforts there is currently no operational CWC product available to users. In the context of the Satellite Application Facility for Land Surface Analysis (LSA-SAF), we have developed an algorithm to produce a global dataset of CWC based on data from the Advanced Very High Resolution Radiometer (AVHRR) sensor on board Meteorological–Operational (MetOp) satellites forming the EUMETSAT Polar System (EPS). CWC reflects the water conditions at the leaf level and information related to canopy structure. An accuracy assessment of the EPS/AVHRR CWC indicated a close agreement with multi-temporal ground data from SMAPVEX16 in Canada and Dahra in Senegal, with RMSE of 0.21 kg m$^{-2}$ and 0.09 kg m$^{-2}$ respectively. Particularly, when the Normalized Difference Infrared Index (NDII) was included the algorithm was better constrained in semi-arid regions and saturation effects were mitigated in dense canopies. An analysis of spatial scale effects shows the mean bias error in CWC retrievals remains below 0.001 kg m$^{-2}$ when spatial resolutions ranging from 20 m to 1 km are considered. The present study further evaluates the consistency of the LSA-SAF product with respect to the Simplified Level 2 Product Prototype Processor (SL2P) product, and demonstrates its applicability at different spatio-temporal resolutions using optical data from MSI/Sentinel-2 and MODIS/Terra & Aqua. Results suggest that the LSA-SAF EPS/AVHRR algorithm is robust, agrees with the CWC dynamics observed in available ground data, and is also applicable to data from other sensors. We conclude that the EPS/AVHRR CWC product is a promising tool for monitoring vegetation water status at regional and global scales.

*Keywords: EUMETSAT Polar System (EPS); AVHRR/MetOp; Canopy Water Content (CWC); Gaussian Process Regression (GPR); MODIS; Sentinel-2*

# 1 Introduction

Water is the abiotic factor which most strongly limits vegetation growth, and vegetation water content is thus an important component of the functioning of terrestrial ecosystems (Nilsen & Orcutt, 1996). It contributes to nutrient and sugar transport, plant metabolism, and buffers plant available water between the uptake via root-soil water and the loss via transpiration (regulated by the stomata). Vegetation water content varies in response to changes in environmental conditions and depends on metabolic activity, plant structure, physiological conditions and hydraulic strategies (Carter *et al*. 1993; Zhang and Zhou, 2015). Additionally, it is a key biophysical variable for agricultural management and yield forecasting (Irmak *et* al., 2000), for water stress and drought assessment (Rossini *et al*., 2013; Rahimzadeh-Bajgiran *et al*., 2012; Asner *et al*., 2016; Martin *et al*., 2018), as well as in flood risk monitoring and wildfire prevention (Riaño *et al*., 2005; Chowdhury *et al*., 2015; Hantson *et* al., 2016; Yebra *et al*, 2013). It is an important component of the hydrological cycle and thereby essential for the carbon and energy cycles, thus affecting ecosystem-climate interactions.

The vegetation water content is usually computed at either leaf or canopy levels; at leaf level it is given by leaf water content ($C_w$), expressed as the leaf fresh weight minus the dry weight per unit leaf area, whereas at canopy level it is the leaf water content times leaf area index (LAI). In this study, canopy water content (CWC) is defined as the amount of water in the leaves per unit ground area. CWC reflects available water conditions at the leaf level and information related to canopy structure.

From a remote sensing point of view, the amount of water present in vegetation can be estimated from sensors working in either the optical or microwave spectral ranges (Moghaddam & Saatchi, 1999; Ceccato *et al*., 2001). When comparing optical and microwave remote sensing approaches to retrieve water content of the vegetation, the former presents an important advantage regarding the maximum possible spatial resolution. In the microwave domain, it can be inferred from active (radar) or passive (radiometer) sensors. In active systems, the backscattering signal (i.e., the ratio of received to transmitted power) is sensitive to vegetation characteristics including its dielectric properties. In passive systems, the attenuation of microwave radiation through vegetation is measured by the Vegetation Optical Depth (VOD) parameter, which is directly related to the amount of water in the vegetation canopy and the above-ground biomass. Sensitivity of microwaves to vegetation water content depends on the frequency of observation, with lower frequencies being more sensitive to deeper canopy vegetation layers (Jackson and Schmugge, 1991, Liu *et al*., 2015, Konings *et al*., 2017). In the optical domain, water absorbs radiation throughout the whole spectrum, but its main absorption features are located at 0.97 µm, 1.20 µm, 1.45 µm, 1.6 µm, 1.94 µm and 2.5 µm (Peñuelas *et al*., 1997; Ceccato *et al*, 2001). Based on the sensitivity of the reflectance to variations in leaf/canopy water content in these wavelengths, various techniques exploiting band ratios and spectral depth produced by the water absorption have been developed (e.g. Ceccato *et al*. 2002; Ullah *et al*., 2014; Yebra *et al*. 2018; Pasqualotto *et al*. 2018). Band ratio indices usually employ one sensitive and one insensitive band to CWC changes (Jackson *et al*., 2004; Trombetti *et al*., 2008). An approach is then to establish a parametric relationship between the index at issue, and corresponding CWC values obtained either from *in situ* measurements or simulated by radiative transfer models (RTMs). In this framework, the Normalized Difference Infrared Index (NDII) (Hardisky *et al*., 1983) has been proposed to detect changes in CWC.

The use of empirical approaches based on spectral indices for retrieving a biophysical variable from *in situ* measurements is a purely statistical technique that lacks generalization since the *in situ* measurements may not cover all the range of variation of the variable of interest. Furthermore, the technique is site-dependent, and requires that measurements are split into calibration and validation sets (Baret & Buis, 2008). On the other hand, RTMs have capacity to generalize and simulate a huge range of vegetation conditions at leaf and canopy levels (Berger *et al*., 2018; García-Haro *et al*, 2018). However, this approach is depending on the correctness of the model in describing physical processes, as well as on the model parameterization. In particular, the PROSPECT (Jacquemoud & Baret, 1990), and the Scattering by Arbitrary Inclined Leaves (SAIL) (Verhoef, 1984) RTMs have been used for estimating vegetation water content for different vegetation types including crops, grasslands, shrublands, and forests, using techniques such as neural networks (NN), look-up tables, and iterative optimization (Weiss *et al*., 2000; Zarco-Tejada *et al*., 2003; Riaño *et al*., 2005; Yebra & Chuvieco, 2009; Zhu *et al*., 2015; Trombetti *et al*., 2008; Djamai *et al*., 2019).

Despite the efforts made by the scientific community, there is not currently operational global-scale CWC product available to users. Some studies have retrieved vegetation water content from multispectral sensors (Zarco-Tejada *et al*., 2003; Jackson *et al*., 2004; Trombetti *et al*., 2008; Djamai *et al*., 2019), but none of them led to an operational product. In this context, Weiss & Baret (2016) developed the Simplified Level 2 Product Prototype Processor (SL2P) for retrieving biophysical variables, including CWC, based on data from the Multispectral Instrument (MSI) onboard the Sentinel-2 satellites. The SL2P is based on the training of a neural network using PROSAIL simulations, but requires downloaded MSI data and execution of a module in the Sentinel Application Platform (SNAP). The SL2P algorithm was validated using MSI simulated imagery (Camacho *et al*., 2013). Results reported in an initial validation of SL2P CWC on MSI images over the SMAPVEX16 site were not entirely satisfactory, with a slope of 0.42 (considerably lower than 1) and a negative bias (−0.37 kg m$^{-2}$) (Djamai *et al.*, 2019). A similar approach was proposed by Campos-Taberner *et al.* (2018) for retrieving global-scale CWC by inverting PROSAIL using random forests and MODerate-resolution Imaging Spectroradiometer (MODIS) data as input. However, no accuracy assessment was provided, and the SL2P product has only been assessed from a small set of measurements, reaching stage 1 of the hierarchy proposed by the Committee on Earth Observation Satellites (CEOS) Working Group on Calibration and Validation (WGCV) Land Product Validation (LPV) sub-group (Fernandes *et al.,* 2014).

The Satellite Application Facility for Land Surface Analysis (LSA-SAF) aims to increase the benefits accrued by the European Organization for the Exploitation of Meteorological Satellites (EUMETSAT) network, specifically to understand and quantify terrestrial processes and land-atmosphere interactions. LSA-SAF contributes to the monitoring of biosphere from space with the generation of products from EUMETSAT satellites on an operational basis for a large number of surface-level variables (Trigo *et al.*, 2011). Today, a suite of global vegetation products from Spinning Enhanced Visible and InfraRed Imager (SEVIRI) onboard the Meteosat Second Generation (MSG) (since 2004) and the Advanced Very High Resolution Radiometer (AVHRR) onboard the EUMETSAT Polar System (EPS) (since 2018) are produced and operationally delivered by LSA-SAF system (García-Haro *et al*., 2018). The suite of MSG and AVHRR vegetation products include LAI, the fractional vegetation cover (FVC), and the fraction of absorbed photosynthetically active radiation (FAPAR), freely disseminated to users through the LSA-SAF website (https://landsaf.ipma.pt) and EUMETCast (https://navigator.eumetsat.int). More recently, the gross primary production (GPP) using MSG (since 2018) data has recently been added to the vegetation products (Martínez *et al*. 2018, 2019). Thus, a climate data record around 15 years is now offered for climate and environmental applications along with the possibility to be enhanced with the availability of new EPS/AVHRR products and other relevant variables for the canopy status, such as GPP from MSG.

In this context, LSA-SAF has developed a new CWC product that will generate global observations from the AVHRR sensor on board Meteorological–Operational satellites (MetOp-A, B and C). The product is targeted for applications such as land carbon fluxes and soil moisture estimation, drought monitoring, and fire risk detection. This paper presents the methodology and aims to demonstrate its feasibility to generate global estimates of CWC. The retrieval algorithm is an improved version of the current LSA-SAF algorithm used for the other operational vegetation products based on AVHRR data (LAI, FVC and FAPAR). The CWC estimation approach leverages the generalization power of PROSAIL, and state-of-the-art machine learning retrieval algorithms. The algorithm encompasses two main features: a more representative parameterization that improves the distributions and co-distribution at leaf level of PROSAIL using a large database of leaf trait measurements extracted from the TRY global database (Kattge *et al*., 2011) and the inclusion of the NDII as a new predictor in order to enhance the sensitivity to water content. In this study, the LSA-SAF algorithm has also been prototyped using MODIS and MSI/Sentinel-2 reflectance with the aim of (1) investigating the applicability of the algorithm to other sensors, (2) assessing the possible influence of scale effects, and (3) evaluating its consistency with respect to the SL2P product. An accuracy assessment of the EPS/AVHRR CWC product was performed using multitemporal ground measurements over two sites located in Canada and Senegal.

## 2 Materials and methods

### 2.1 Ground observations of CWC

Two study areas were considered for assessing the quality of CWC retrievals. The first one was located in Manitoba, southern Canada (49.90 N, 97.14 W, Figure 1). The site has been extensively used for development, calibration and validation of products from the NASA's Soil Moisture Active Passive (SMAP) mission (Entekhabi *et al*., 2010). The study area contains crop fields consisting mainly of soybean, canola, corn, wheat and oat. In this work we will use the intensive ground-based CWC measurements collected during the SMAP Validation Experiment in 2016 (SMAPVEX16), which took place during two 2-week periods: from 8–20 June and 10–22 July, 2016 (Cosh *et al*., 2019). Each field was sampled at least once over the two-week period. For canola, wheat, oat and black bean fields, destructive measurements of all aboveground biomass were obtained within three 0.5 m × 0.5 m squares per field. These three squares were distributed across the field to be representative of the sampled field as a whole. For corn and soybeans fields, five plants along two rows were collected and the biomass of these plants was then scaled to a unit ground area (m$^2$) using the density of crop. For soybean, canola, corn and black bean a distinction was made between leaves and stems, whereas for wheat and oat water from only total aboveground biomass was measured. All biomass samples were weighted in the field to get fresh weight. The biomass samples were then dried and weighted again to get dry weight. CWC was obtained by subtracting the dry from the fresh weight. For further description of SMAPVEX16 data set, we refer to Cosh *et al*., (2019). SMAPVEX16 Manitoba *in situ* vegetation data was downloaded from the National Snow and Ice Data Center Distributed Active Archive Center (McNairn *et al*., 2018).

The second study area was located in Senegal, northeast of the city of Dahra (15°24′10″N, 15°25′56″W, elevation 40 m, Figure 1) belonging to the Sahelian ecoclimatic zone. The Dahra field site has a ~ 3 months long growing season following the rainy season (July-October), with LAI generally ranging between 0 and 2 m$^2$ m$^{-2}$ (Fensholt *et al*., 2004). There is a dominance of annual grasses (e.g., *Schoenefeldia gracilis, Digitaria gayana* and *Dactyloctenium aegypticum*) (Mbow *et al*., 2013), trees and shrubs (e.g., *Acacia Senegal* and *Balanites aegyptiaca*) relatively sparse (~ 3 % of the land cover) (Rasmussen *et al*., 2011). The average tree height is 5.2 m and the peak height of the herbaceous layer is 0.7 m (Tagesson *et al*., 2015). Note that the Dahra field site was selected to be representative of the spatial resolution of the satellite imagery used in the study (i.e., a ~500 m MODIS pixel), even though an acacia plantation has been established about 500 meters south of the field site. A thorough description of the Dahra field site is given in Tagesson *et al*. (2015). At the Dahra field site, total above ground herbaceous biomass (kg m$^{-2}$) was sampled every 10 days during the growing seasons 2008-2018 at 28 one m$^2$ plots located along two ~1 km long transects (Mbow *et al*., 2013). All above ground green vegetation matter was collected and weighed in the field to get the fresh weight. The collected biomass was oven-dried and then weighted to get dry weight. The CWC was obtained by subtracting the dry from the fresh weight. As the tree canopy cover is just ~3% of the ground surface and all herbaceous plants were destructively harvested (Tagesson *et al*., 2015), upscaling of leaf water content to the canopy level using LAI was not required.

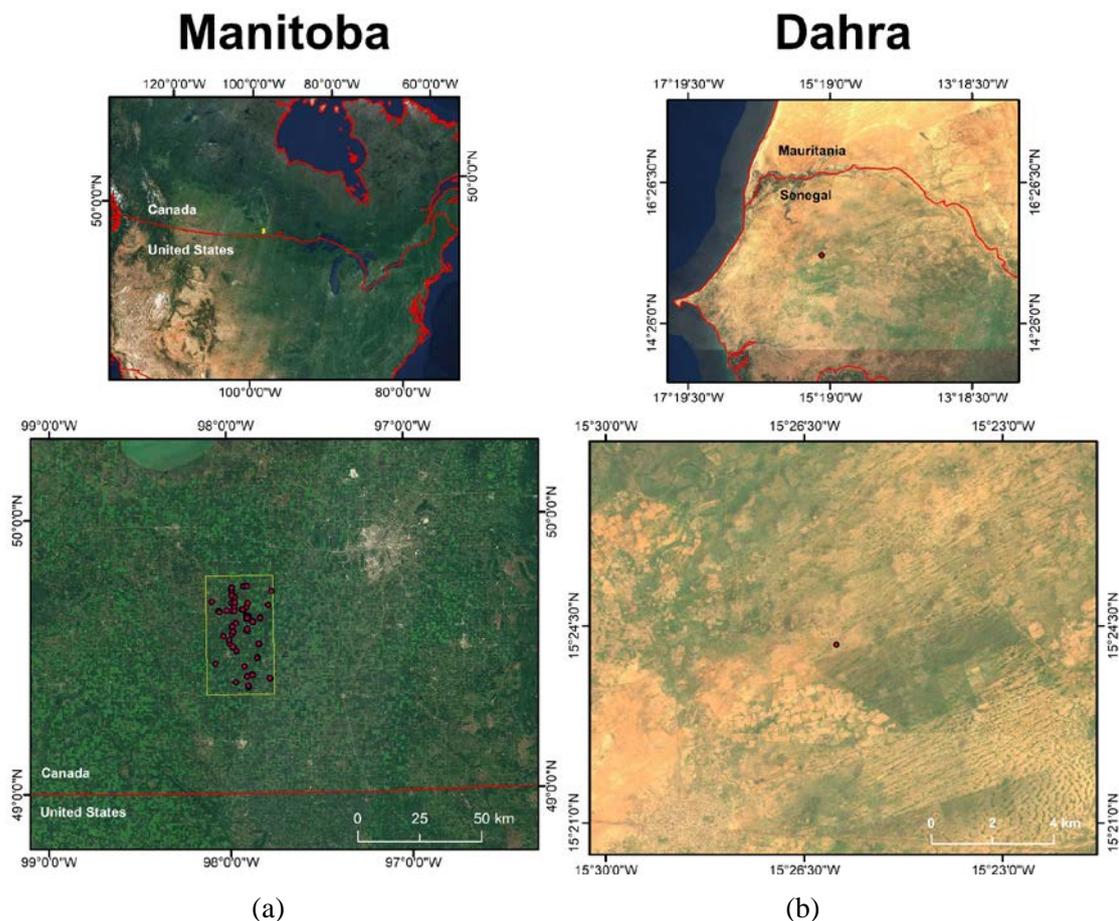

Figure 1. Location of the study areas. (a): SMAPVEX16 Manitoba site, and (b): Dahra site. Location of ground measurements are represented by red circles.

## 2.2 Retrieval algorithm

The LSA-SAF disseminates land surface albedo from measurements acquired by the AVHRR sensor. The albedo is computed using a semi empirical bidirectional reflectance distribution function (BRDF) based on a decomposition of the bi-directional reflectance factor into a number of kernel functions ($k_0$, $k_1$, $k_2$) associated with dominant light scattering processes (Roujean *et al*., 1992). The input of the CWC retrieval algorithm is the normalized spectral reflectance factor ($k_0$) in three AVHRR bands at 0.63 µm, 0.87 µm, and 1.61 µm (Figure 2). Detailed information of the LSA-SAF albedo product is given in Geiger *et al*., (2018).

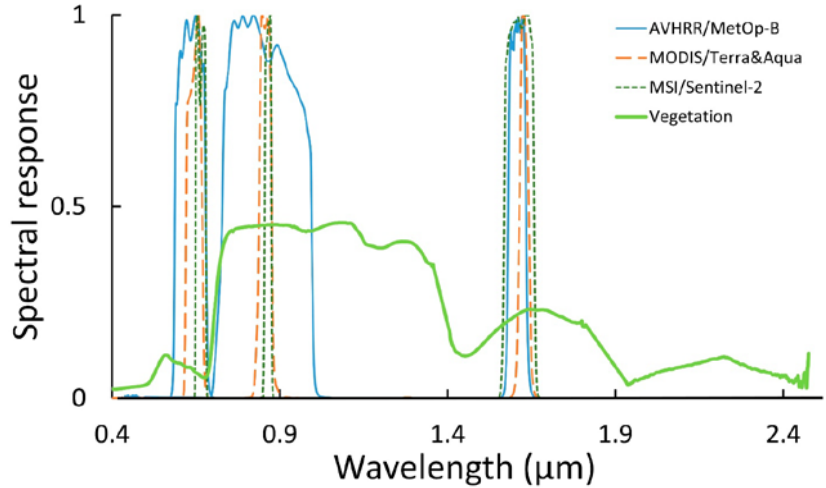

Figure 2. Spectral response functions of AVHRR optical channels (1, 2 and 3), centered at 0.630, 0.865 and 1.610 µm, respectively, onboard MetOp-A and MetOp-B, as well as the corresponding MODIS and MSI/Sentinel-2 spectral bands. The vegetation spectrum shown is for illustration purpose only.

The CWC retrieval algorithm uses a machine learning regression method that learns the relationship between the observed reflectance and the chosen canopy parameters. The model is trained on simulations and applied afterwards on real reflectances. We used Gaussian processes for regression (GPR), which have shown excellent accuracy capabilities to deal with multivariate and nonlinear feature relations (Camps-Valls *et al*, 2016; 2019). In particular, the approach is an improved version of the current algorithm used for the generation of the EPS/AVHRR vegetation variables as it relies on a hybrid approach based on multi-output GPR ($GPR_{multi}$) trained on PROSAIL simulations (García-Haro *et al*., 2018).

The retrieval of vegetation biophysical variables in any RTM inversion scheme has a wide range of possible solutions, and the use of improved prior information has been recognized as an appropriate way to constrain them (Combal *et al*., 2003). This prior information is usually extracted from the literature, provided by an expert, or inferred from a compilation of experimental data. In our case, we combined information available in the literature and a global repository for plant trait data to optimize our prior information and minimize ambiguities in the solutions. The selected plant trait database (TRY) is the biggest available at the moment with an unprecedented spatial and climatological coverage. It encompasses a repository of plant trait data with more than 11 million records in version 5, the number of which is continuously growing. The PROSAIL parameterization at the leaf level is based on the distribution functions extracted from the TRY database by means of a kernel density estimator (KDE) (Parzen, 1962). The TRY database includes leaf water content ($C_w$ in kg m$^{-2}$), leaf chlorophyll ($C_{ab}$ in kg m$^{-2}$), and leaf dry matter ($C_{dm}$ in kg m$^{-2}$) at global scale for a wide range of species (see distributions of $C_w$ in Figure 3). In the TRY database, traits are not abundance-weighted with respect to natural occurrence (Kattge *et al.*, 2011), and the corresponding distributions are therefore biased. To overcome this bias, we weighed the distributions by plant functional type (PFT) occurrence according to the MODIS land cover product (Friedl *et al.*, 2010).

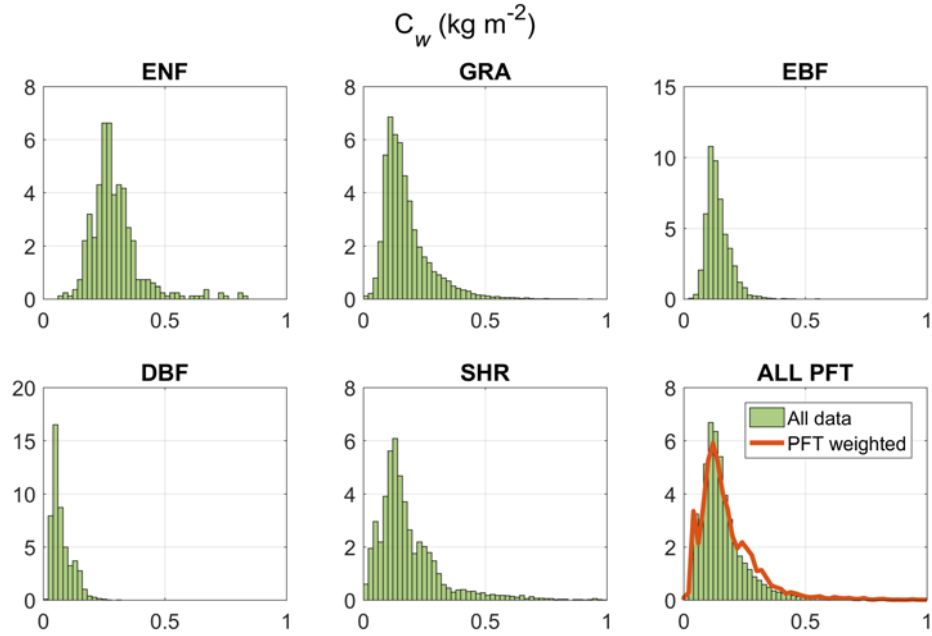

Figure 3: Histograms of leaf water content ($C_w$) measured in the TRY database per plant functional type (PFT) (ENF: evergreen needleleaf forest; GRA: grassland; EBF: evergreen broadleaf forest; DBF: deciduous broadleaf forest; SHR: shrubland).

In addition, in order to capture the tradeoff and dependencies among $C_w$, $C_m$, and $C_{ab}$, a multivariate Gaussian copula function (Žežula, 2009) was used to obtain the samples used for distributions also retaining the correlation among leaf traits (see Figure S1 in supplementary material). Remaining leaf parameters (the mesophyll structural parameter (N), and leaf brown pigment ($C_{bp}$)) distributions, and the canopy parameters (LAI, average leaf angle (ALA), and the hot-spot parameter (Hotspot)) used in PROSAIL, were set up according to similar retrieval studies at global scale (Baret *et al.*, 2007; García-Haro *et al.*, 2018; Campos-Taberner *et al.*, 2018) (Table 1).

To account for heterogeneity at the landscape level each surface was assumed to be a mixture of the spectral contributions of vegetation (*vCover*) and bare area (*1-vCover*) fractions. Time series of EPS/AVHRR FVC images were used to identify spectra of bare areas representing the high variability in global soil conditions (García-Haro *et al.* 2018). The geometry of the sun-sensor was characterized by the solar and view zenith angles, and the relative azimuth angle. In our case they corresponded to illumination and observation zenith angles of 0°. The parameters at leaf and canopy levels are outlined in Table 1.

Table 1. Ranges and distributions of the PROSAIL parameters adopted in the EPS/AVHRR retrieval chain. (*) A 5% of the 3000 spectral samples representative of pure background (vCover=0) were included to account for bare areas. KDE refers to a fitted kernel distribution function.

|  | Parameter | Min | Max | Mean | Std | Type |
|---|---|---|---|---|---|---|
| Canopy | LAI (m$^2$/ m$^2$) | 0 | 8 | 3.5 | 4 | Gaussian |
|  | ALA (°) | 35 | 80 | 62 | 12 | Gaussian |
|  | Hotspot | 0.1 | 0.5 | 0.2 | 0.2 | Gaussian |
|  | vCover | 0.3 | 1 | 0.99 | 0.2 | Truncated gaussian(*) |
| Leaf | N | 1.2 | 2.2 | 1.5 | 0.3 | Gaussian |
|  | $C_{ab}$ (kg m$^{-2}$) | - | - | - | - | KDE |
|  | $C_{ar}$ (µg·cm$^{-2}$) | 0.6 | 16 | 5 | 7 | Gaussian |
|  | $C_{dm}$ (kg m$^{-2}$) | - | - | - | - | KDE |
|  | $C_w$ (kg m$^{-2}$) | - | - | - | - | KDE |
|  | $C_{bp}$ | 0 | 0 | 0 | 0 | - |
| Soil | $\beta_s$ | 0.1 | 1 | 0.8 | 0.6 | Gaussian |

The PROSAIL model was executed in forward mode (reflectance at 0.001 µm resolution in the range of 0.4 µm to 2.5 µm) for simulating samples of biophysical variables to be retrieved. We tested the robustness of the method against the number of training samples and stable results were obtained when more than 2500 samples were considered (García-Haro et al., 2018). Therefore 3000 samples were finally used to train our model. The top-of-canopy reflectance (TOC-R) in the AVHRR channels was obtained filtering with the spectral response of the AVHRR instrument onboard MetOp (see Figure 2). Subsequently, a moderate level of zero-mean Gaussian noise $\mathcal{N}(0, \sigma^2)$, with standard deviation σ=0.015, was added to simulated reflectances, to enhance the stability of solutions and reduce overfitting (García-Haro *et al*. 2018).

On the basis of the outcome of the PROSAIL simulations (see Section 3), the three AVHRR channels jointly with the NDII (NDII = $(\rho_{0.8} - \rho_{1.6})/(\rho_{0.8} + \rho_{1.6})$), here calculated from channels 2 (0.865 µm) and 3 (1.610 µm), were used as predictors for making the retrieval model more sensitive to changes in vegetation water content and minimize the sensitivity to other leaf and canopy variables and soil background. The NDII index has proven to improve the estimates of biophysical variables such as fuel moisture content using RTM based approaches (Yebra *et al*., 2009, 2018).

One key feature of the GPR$_{multi}$ model is its method of inferring model parameters. GPR works by defining a covariance (kernel) function that estimates similarities among samples (reflectances). The covariance is explicitly defined and normally contains a set of kernel hyperparameters to be adjusted. We followed the same approach for inference in the multi-output case as the one in (García-Haro *et al* 2018). After some initial comparisons to other machine learning methods, we observed that the proposed multi-output GPR (GPR$_{multi}$) provided a slight increase in estimation accuracy as compared with plain GPR, neural networks (NN) and Kernel Ridge Regression (see Table S1 in Supplementary material). Furthermore, the method offered some features that make it advantageous in our setting:

(i) it provides pixel-wise confidence intervals for the predictions (Campos-Taberner *et al*., 2016) also allowing for uncertainty propagation under a solid probabilistic treatment (Camps-Valls *et al*, 2019; Johnson *et al*., 2019), (ii) it does not require training each biophysical variable separately but optimizes the similarities using the complete data set, enhancing thus the consistency, and outperforming other multi-output techniques such as NN (García-Haro *et al*., 2018).

Using the pixel-wise confidence intervals, the algorithm also provides a confidence interval for the CWC estimate, u(CWC), as described in García-Haro *et al*. (2018). It is quantified considering the uncertainty propagation of the inaccuracies associated with the BRDF parameters ($\sigma_{k_0}$) and the standard deviation of GPR prediction ($\sigma_{GPR}$). It should be noted that the $k_0$ covariance matrix assumes that distribution of errors are Gaussian and mutually uncorrelated.

## 2.3 Application of the algorithm to other sensors

The LSA-SAF algorithm for CWC estimation was also prototyped using MSI/Sentinel-2 and MODIS reflectance in order to evaluate its versatility and potential application to optical sensors with different spectral and spatio-temporal characteristics. For both datasets, we used their corresponding spectral bands in the red, NIR and SWIR channels. PROSAIL simulations were performed taking into account the spectral bands responses functions of the different sensors.

### 2.3.1 MSI/Sentinel-2 data

The European Space Agency grants free access to Copernicus Sentinel-2 satellite constellation (Drusch *et al*., 2012) data since the end of 2015 to date. The Sentinel-2 mission comprises two satellites (Sentinel-2A, and -2B) carrying the MSI, which measures top-of-atmosphere reflectances (Level 1C) in 13 channels distributed in the visible (VIS), the NIR, and the SWIR spectrum. The Level-1C products were downloaded from the Sentinels Scientific Data Hub (https://scihub.copernicus.eu/) and subsequently atmospherically corrected by means of executing the Sen2cor (release 2.5.5) processor in SNAP version 6.0, thus obtaining Level-2A TOC-R. For consistency with the coarse resolution products, we used as input MSI/Sentinel-2 channels 4, 8A and 11, centered at about 0.665 μm, 0.865 μm and 1.610 μm (Figure 2). For the SMAPVEX16 area, seven images corresponding to two MSI/Sentinel-2 granule tiles (T14UNA and T14UNV) were used while at the Dahra field site 32 images corresponding to granule tile T28PDC were considered for years 2016-2018. The Quality Scene Classification layer produced by Sen2Cor was used to mask cloudy/shadowed pixels.

### 2.3.2 MODIS/Terra & Aqua BRDF parameters

The Bidirectional Reflectance Distribution Function (BRDF)/Albedo Model Parameters product (MCD43A1) is generated daily at a 500 m resolution through composite periods of 16 days. The product uses all the acquisitions from the MODIS sensors onboard both NASA's Terra and Aqua satellites from the retrieval period. The 9th day of the 16-day MCD43A1 retrieval period was chosen to coincide with the center of the 20-day composite of the EPS/AVHRR CWC product. The images were therefore downloaded with a 10-day frequency for tiles Th16v07 (Dahra site) between January 2008 and December 2018, and for tile h11v04 (SMAPVEX16 site) for the period from March to November 2016. In this study, we used the $k_0$ (isotropic) BRDF parameter in channels 1, 2 and 6, centered at about 0.665 μm (red), 0.865 μm (NIR) and 1.609 μm (SWIR 1). This is a TOC-R directionally normalized to illumination and observation zenith angles of 0°.

## 2.4 Intercomparison with the Sentinel-2 Level-2B product prototype processor (SL2P)

The EPS/AVHRR CWC was intercompared over the SMAPVEX16 and Dahra field sites with CWC estimated by applying the SL2P module on the MSI/Sentinel-2 Level-2A images within the SNAP software. The SL2P module network retrieves CWC using as inputs 8 MSI/Sentinel-2 bands (B3, B4, B5, B6, B7, B8a, B11, and B12) and the cosine of the sun zenith angle, view zenith angle, and relative azimuth angle (Weiss & Baret, 2016).

## 2.5 Data analysis

The CWC estimates based on MSI and MODIS data were resampled to the same sinusoidal grid as the EPS/AVHRR and then all valid observations within each EPS/AVHRR pixel were averaged. A no-data value was assigned if more than 30 % of the pixels were unreliably calculated. In order to perform the assessment of satellite products, we applied a cubic-splines interpolation method to the satellite data. The interpolated satellite observations were associated with *in situ* measurements if a minimum of *N*=4 satellite observations were within an interval of ± 15 days centered on the *in situ* measurement.

In the SMAPVEX16 area, the accuracy was assessed against ground observations, up-scaled with high resolution MSI imagery for a proper spatial representativeness at the kilometric pixel scale. In the more homogeneous Dahra site, the temporal consistency of the EPS/AVHRR CWC product (as well as the CWC derived using MSI and MODIS sensors) was assessed using time series of destructive measurements. In order to undertake the assessment, the EPS/AVHRR product for several representative areas was compared with the aggregated values of 10 m resolution maps obtained using pure statistical methods based on a transfer function (TF) recommended by CEOS WGCV LPV (Morisette *et al.*, 2006; Fernandes *et al.*, 2014). Random forest regression (Breiman, 2001) was used to derive ground based maps through a TF linking field measurements of water content with interpolated MSI/Sentinel-2 surface reflectance. These maps were used to evaluate the EPS/AVHRR coarse resolution product in a multi-temporal manner.

Finally, in order to quantify the possible impact of scale effects on the LSA-SAF algorithm, in Section 3.7 we compare the results of its application at two spatial resolutions: a high resolution LSA-SAF MSI CWC map degraded to EPS/AVHRR resolution and a product derived at coarse EPS/AVHRR resolution using degraded MSI/Sentinel-2 reflectance.

The following metrics were used to assess the differences between observations and estimates: Root Mean Square Error (RMSE), Relative RMSE to range $\left(\text{rRMSE}(\%) = \frac{\text{RMSE}}{(\max(y_i) - \min(y_i))} \times 100\right)$, bias and coefficient of determination ($r^2$).

# 3 Results

## 3.1 The EPS/AVHRR CWC product

Using the LSA-SAF CWC retrieval algorithm, global maps of CWC were produced for the period between 21$^{st}$ January 2015 and 31$^{st}$ December 2018 in sinusoidal projection every 10 days, with a resolution of about 1.1km × 1.1km. In order to reduce the sensitivity to outliers and to extended periods of missing data due to persistent cloudiness, the LSA-SAF albedo product uses a temporal weight function through composite periods of the last 20 days with a characteristic temporal scale (full width at half mean) of 10 days (Geiger *et al.*, 2018). For example, the day of production March 25$^{th}$ corresponds to the period March 6$^{th}$-25$^{th}$, and observations collected during the composite window are assigned different weights: 100% for March 25$^{th}$, 50% for March 15$^{th}$ and only 10% for March 6$^{th}$.

Each EPS/AVHRR CWC product is distributed in Hierarchical Data Format 5 (HDF5) and contains 3 relevant layers: the product itself, its uncertainty estimate, and a quality flag layer with information about the retrieval conditions and presence of snow/ice and water. Examples of CWC at two periods, spring and winter, are shown in Figure 4. The results are coherent with CWC global expected patterns, showing higher values in spring for most of the northern hemisphere. The product presents a good spatial coverage, except over northern latitudes in winter time. The larger uncertainties (>0.4 kg m$^{-2}$) are usually found in regions with poor retrieval conditions (due to illumination angles and snow/cloud contamination) and reduced sensitivity associated with saturation effects. A preliminary qualitative assessment of the products indicates that the results are realistic and artifact-free. Additionally, they are spatially and temporally consistent with other biophysical variables such as LAI, as it is further detailed in Section 4.2

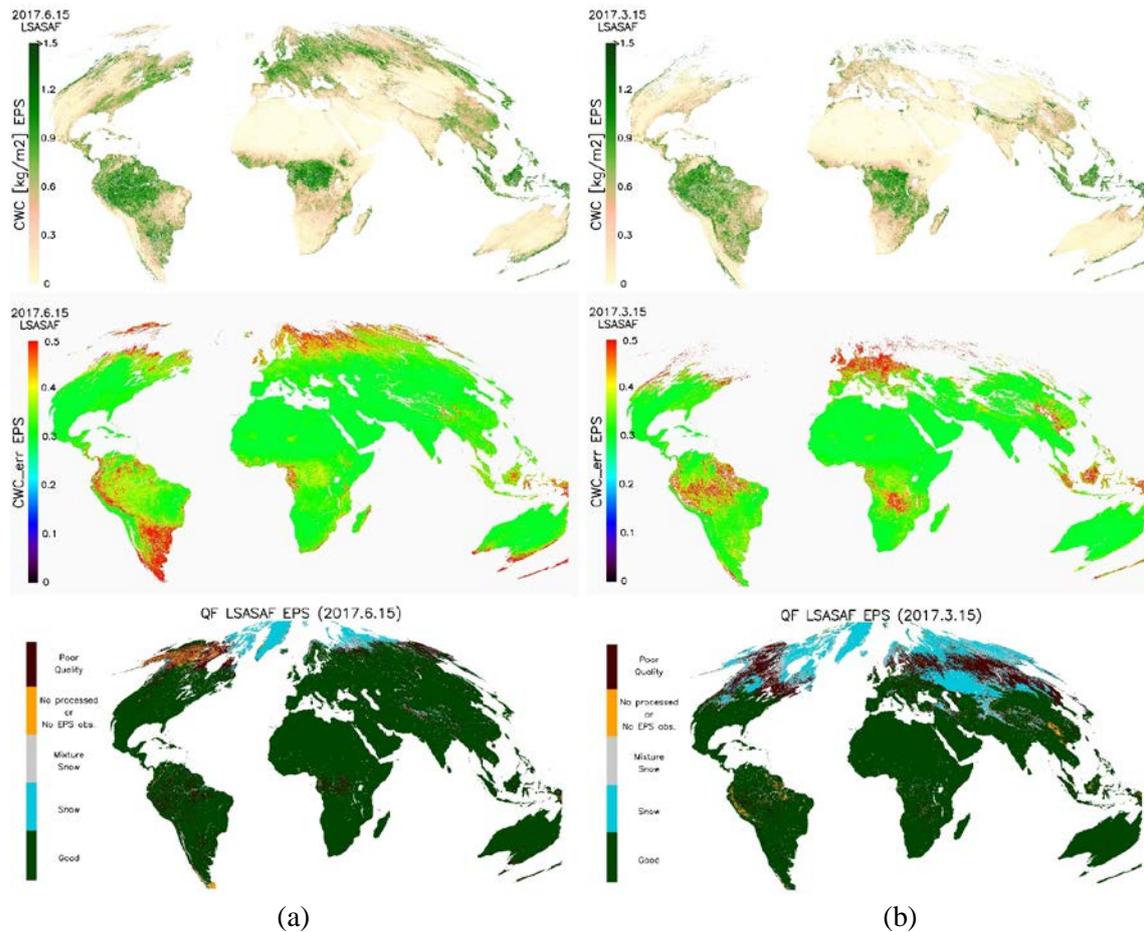

(a)                  (b)

Figure 4. CWC (top), uncertainty estimates (middle) and quality flag information (bottom) of the EPS/AVHRR CWC produced at two periods: (a) 26$^{th}$ May- June 15$^{th}$ 2017; (b) 24$^{th}$ February- March 15$^{th}$ 2017.

## 3.2 Sensitivity assessment of the impact of CWC on reflectance

Similar levels of CWC may generate different spectral signatures depending on LAI values (Figure 5). We can observe that AVHRR channels 1 and 2 (in red and NIR regions, respectively) show small variations to leaf water content, while channel 3 (SWIR region) shows a high variability.

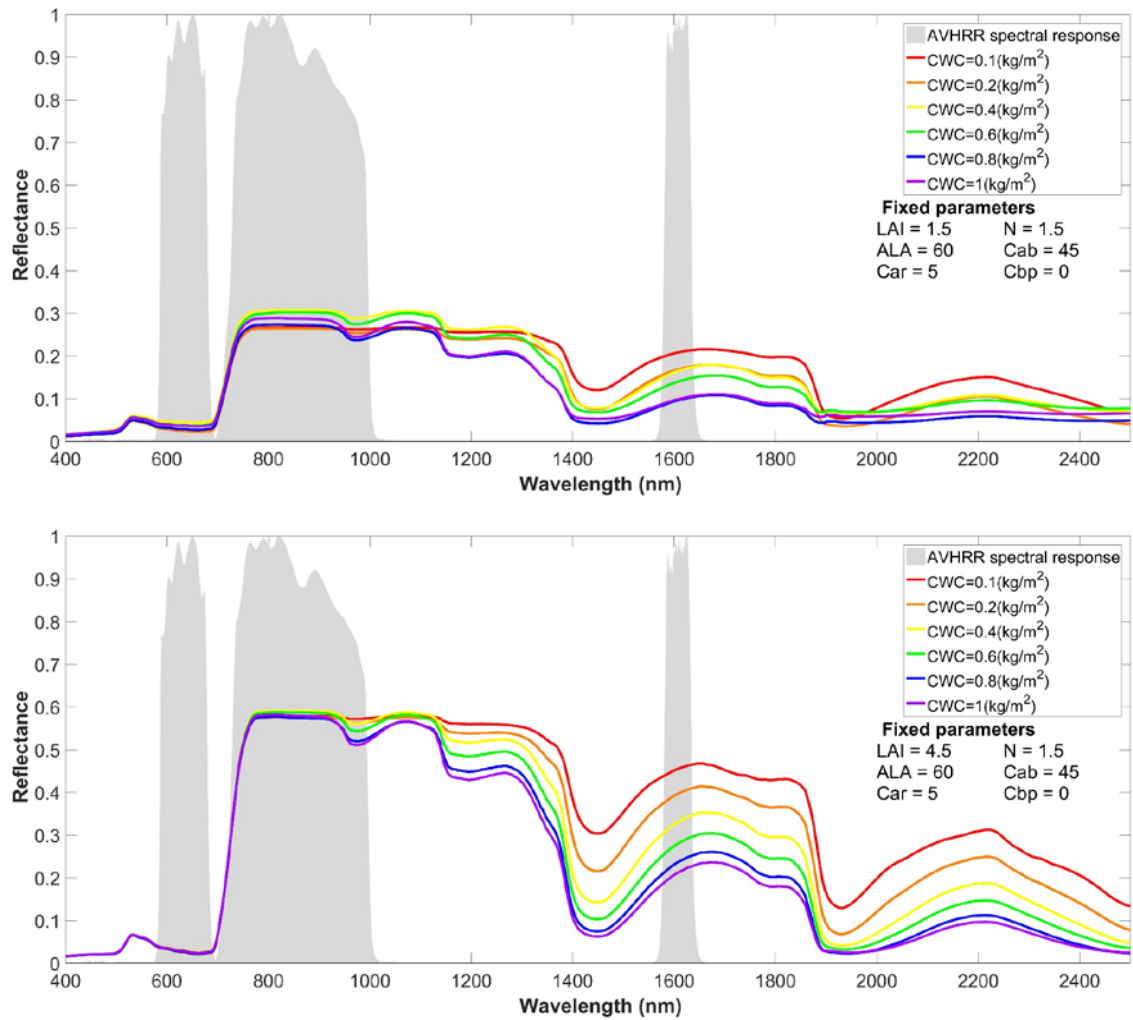

Figure 5. Spectral signatures obtained with PROSAIL at a fixed LAI representing an intermediate canopy, LAI=1.5 (top), and a dense canopy, LAI=4.5 (bottom). In both cases, identical CWC values ranging from 0.1 to 1 are considered. The spectral responses of the three AVHRR bands is also shown in gray.

PROSAIL simulations in Figure 6(a) demonstrate that CWC considerably affects reflectance at SWIR wavelengths (channel 3). However, the variability of background reflectance causes a substantial range in SWIR values for sparsely vegetated surfaces (see Figure S2 in supplementary material). In turn, NDII shows a very good sensitivity to CWC and considerably reduces the confounding effects of soil on CWC (see Figure 6(b)). The SWIR reflectance shows a certain saturation effect when vegetation water content is larger than about 2 kg m$^{-2}$. Large errors are thus expected for CWC in dense canopies. Although NDII was specifically designed for water content, simulations in Figure 6(c) show a strong relationship between NDII and LAI, with moderate saturation effects, suggesting that NDII may be a valuable predictor of LAI. Since our aim is to produce consistent retrievals of CWC and the other EPS/AVHRR based biophysical variables (i.e. LAI, FVC and FAPAR), the NDII has been included as a predictor in the new LSA-SAF algorithm for a joint retrieval of the four variables using an identical algorithm. Further insight of the positive impact of using NDII for the retrievals is provided in Section 3.3.

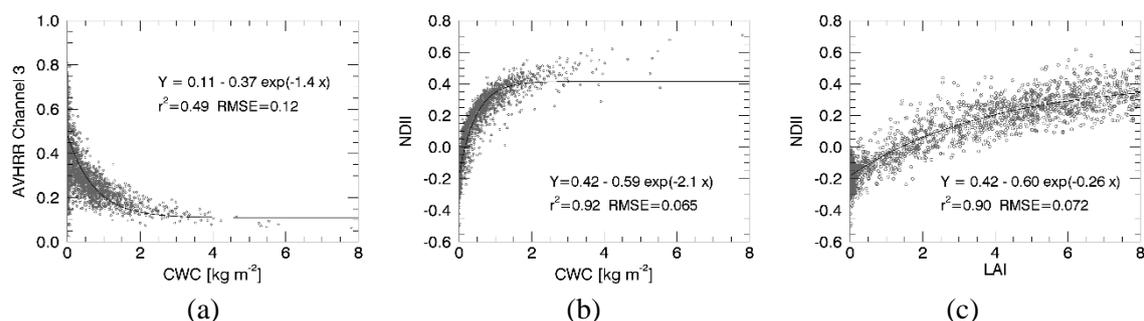

| (a) | (b) | (c) |

Figure 6: Relationship between CWC and PROSAIL simulations. (a): spectral reflectance in AVHRR channel 3 *vs*. CWC; (b) NDII *vs*. CWC; (c) NDII *vs*. LAI. A fit to a parametric (exponential) model is also included.

## 3.3 Influence of the use of NDII

In order to investigate the potential benefits of using NDII as predictor, temporal profiles of EPS/AVHRR products were generated during a four-year period over the BELMANIP2.1 network (Baret *et al*., 2006). This network originally includes 445 sites specially selected for intercomparison of land biophysical products but was further complemented with locations coming from the DIRECT 2.0 OLIVE network (Camacho *et al*., 2017).

The inclusion of NDII as predictor produces almost negligible differences in LAI, FVC and FAPAR estimates for the majority of sites (see Figure 7). In general, CWC presents also slight differences for the majority of sites during the entire cycle. However, for a few sites over semi-arid regions and shrublands, some anomalous patterns are identified when NDII is not included as predictor (Figure 8). The examples in Figure 8 over shrublands cover reveal anomalous peaks between the summer decline and the onset of growth whereas LAI presents realistic (very low) values. Note that in the case of Senegal (close to the Dahra field site), very low values are expected during the dry season from November to June.

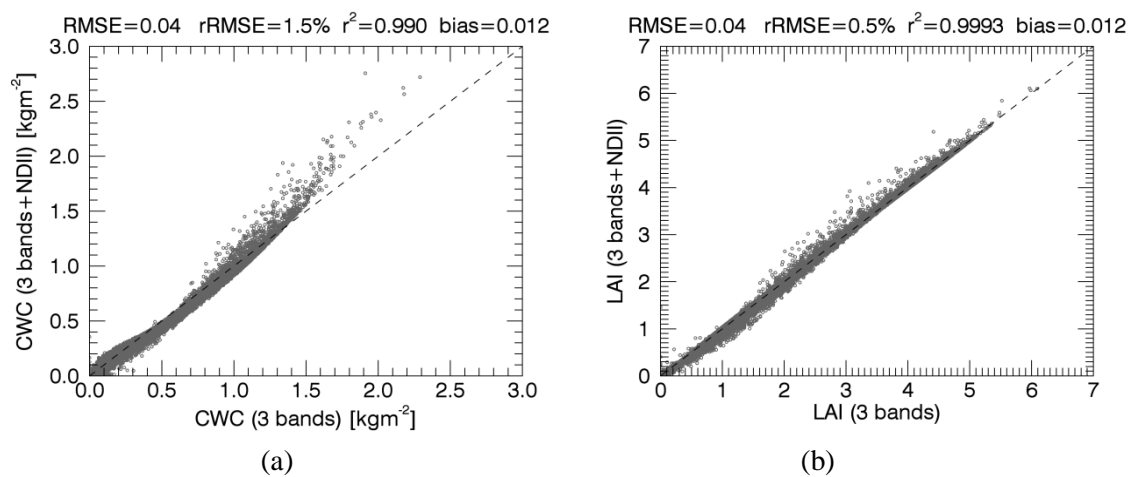

| (a) | (b) |

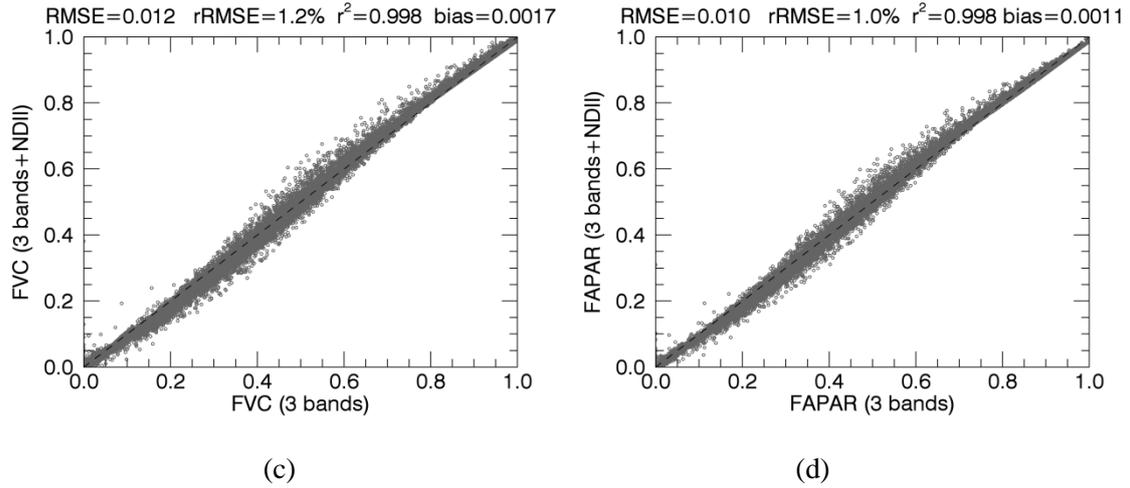

Figure 7. Influence of the use of NDII on retrieved biophysical variables (a) CWC, (b) LAI, (c) FVC, and (d) FAPAR. Scatterplots correspond to all BELMANIP2.1 sites over a four-year period (2015-2018).

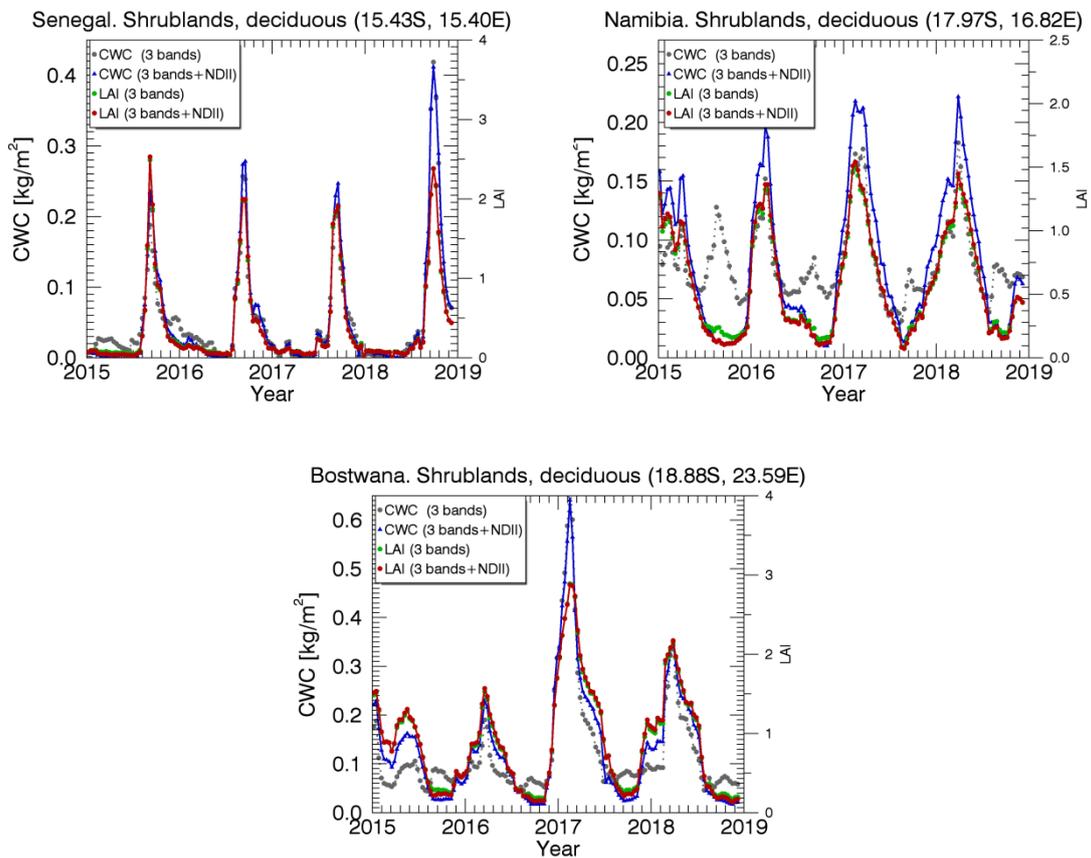

Figure 8. Temporal profiles of CWC and LAI derived from EPS/AVHRR with and without using NDII as predictor over 3 homogeneous BELMANIP 2.1 sites.

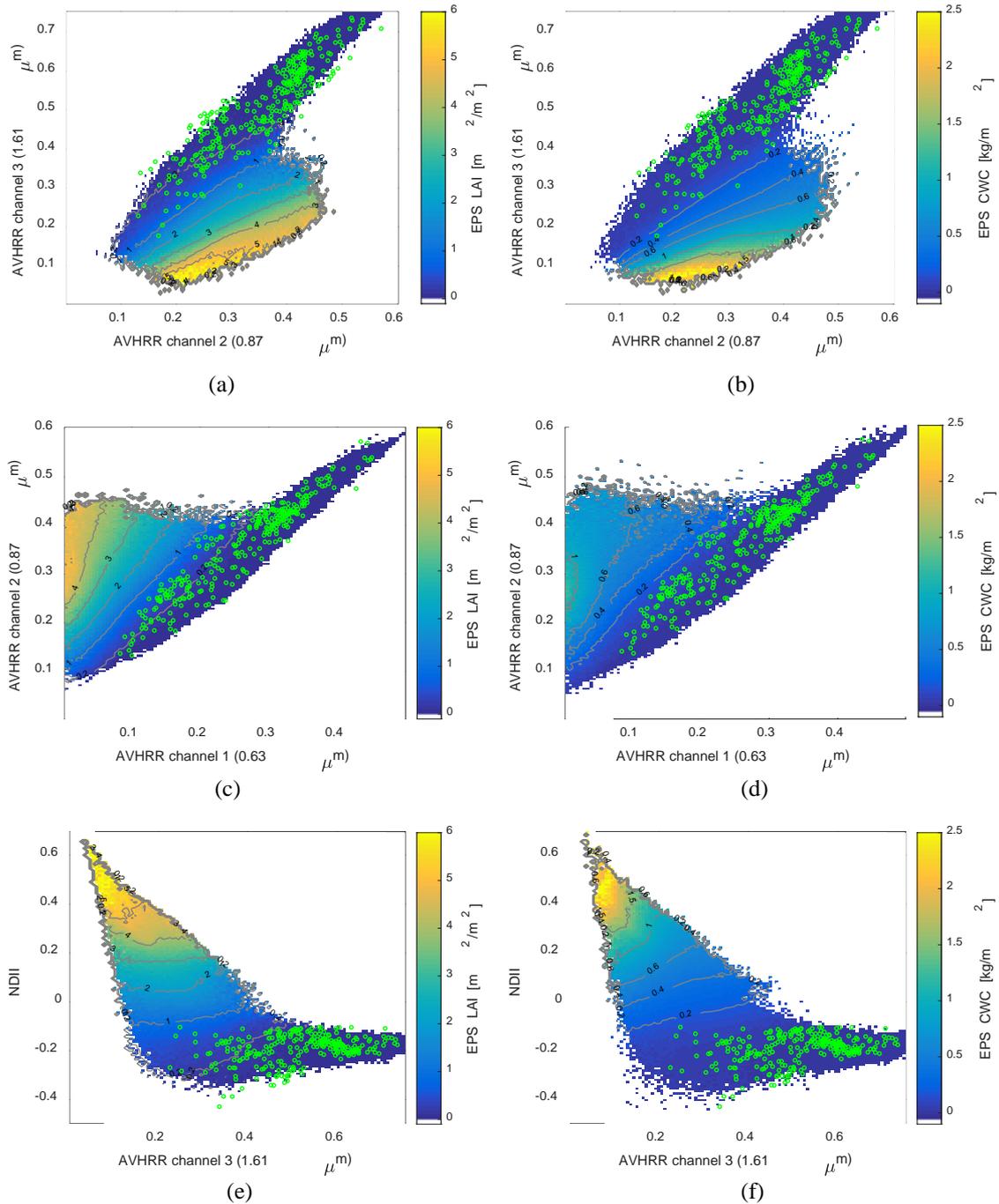

Figure 9. Projection of the LAI (left) and CWC (right) EPS/AVHRR global estimates, onto bidimensional AVHRR/MetOp-B features: channels 1 and 2 (Top); channel 2 and 3 (middle); channel 3 and NDII (Bottom). The isolines of constant LAI or CWC are drawn. The spectral responses of bare areas used as input in PROSAIL are represented by green circles.

Since we wanted to better understand the applicability of the EPS/AVHRR inputs for the estimation of CWC and LAI, further analyses were carried out. As an example, the scatterplots in Figure 9 illustrate how the mean value of one retrieved variable varies in the feature space of two predictive variables for all possible realizations. The high sensitivities of the channels 1 and 2 to LAI make the red near-infrared feature space useful to retrieve canopy biophysical variables such as LAI (Figure 9(a)). However, because of the high water absorption sensitivity in these wavelengths as compared to SWIR spectral bands, these channels are limited in discriminating high water content values exceeding 1.0 kg m$^{-2}$ (Figure 9(b)). A broader variation of CWC is found in the feature space of channels 2 and 3, with values ranging from 0 to 2 kg m$^{-2}$ (Figure 9(d)). It is also noteworthy that the CWC gradient is dominated by channel 3 (SWIR reflectance) due to the high sensitivity of this channel to water variations. In the feature space of channel 3 and NDII (Figure 9(e) and (f)), EPS/AVHRR spectral data lie in a triangle shaped region with the top vertex corresponding to the highest LAI and NDII values. These results confirm the greater sensitivity of this index to variations of these biophysical variables and justify its inclusion in the predictive model.

### 3.4 Accuracy assessment in the SMAPVEX16 area

Temporal dynamics of the EPS/AVHRR product were assessed for five regions of interest (ROIs), each covering an area of 3×3 EPS/AVHRR pixels centered on locations of fields representative of the dominant crops sampled during SMAPVEX16 (Figure 10).

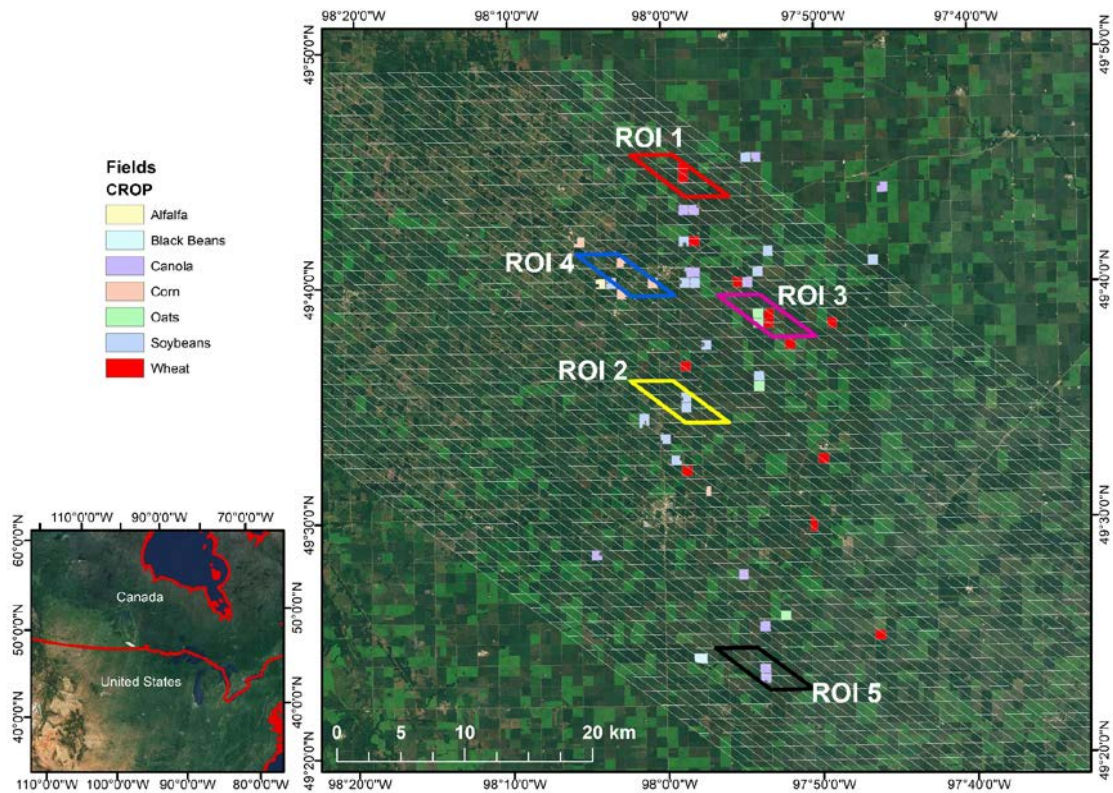

Figure 10. Location of the SMAPVEX16 study area (white 130×130 pixel EPS/AVHRR grid). The right panel provides a zoom over the study area showing the EPS/AVHRR grid (white), sampled fields, and selected ROIs. (Basemap: EOX).

Figure 11 shows the comparison between the EPS/AVHRR CWC and the aggregated for the ROI ground-based TF values of total CWC (CWC$_{leaves+stems}$) and CWC from leafs only (CWC$_{leaves}$). Additionally, aggregated values of CWC estimates derived using the LSA-SAF algorithm on MSI/Sentinel-2 data and MODIS are also shown. In general, the EPS/AVHRR CWC agrees well with TF CWC$_{leaves}$ but does not represent TF values from leaves and stems. The CWC estimates derived from EPS/AVHRR, MSI/Sentinel-2 and MODIS data agree reasonably well, which is indicative of the robustness of the presented algorithm. The EPS/AVHRR and MODIS temporal curves are highly consistent at all sites during the entire year. However, the EPS/AVHRR presents a negative bias during the growing season with respect to MODIS and MSI/Sentinel-2 estimates at wheat and oat sites.

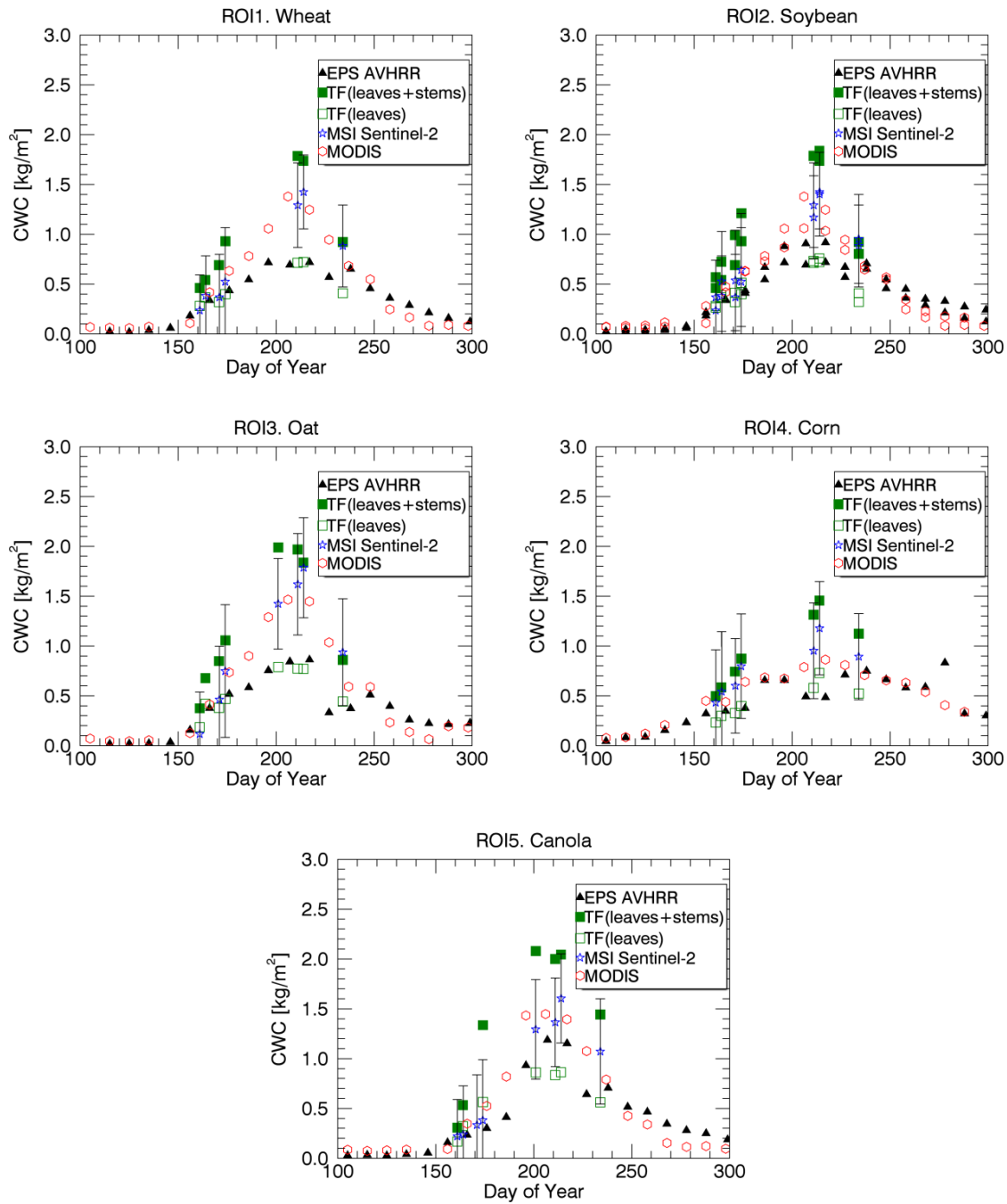

Figure 11. Temporal comparison of CWC products among EPS/AVHRR, MSI/Sentinel-2 and MODIS for 5 representative areas, jointly with the TF over representative areas during 2016. Vertical bars display standard deviation of MSI observations. For the sake of visualization, only the error bars associated with the MSI observations are displayed, which represent the spatial variability within the corresponding ROI.

In order to quantitatively assess the differences between the EPS/AVHRR product and ground based TF maps, EPS/AVHRR CWC observations and TF values were matched for the corresponding dates during 2016. From the scatter plot (Figure 12) it is observed that EPS/AVHRR observations are closer to $CWC_{leaves}$ (RMSE=0.19 kg m$^{-2}$, rRMSE=21%, $r^2$=0.78, bias=0.07 kg m$^{-2}$, n=36 samples), although a slight overestimation for high water content values is also noticeable. In relation to TF ($CWC_{leaves+stems}$), EPS/AVHRR values show a good correlation ($r^2$=0.81) but a clear systematic underestimation (RMSE=0.66 kg m$^{-2}$, rRMSE=31%, bias= −0.56 kg m$^{-2}$, n=36 samples) (Figure 12). These results suggest that EPS/AVHRR estimates are primarily sensitive to top of canopy foliage elements, being rather insensitive to water from the lower parts of the plant.

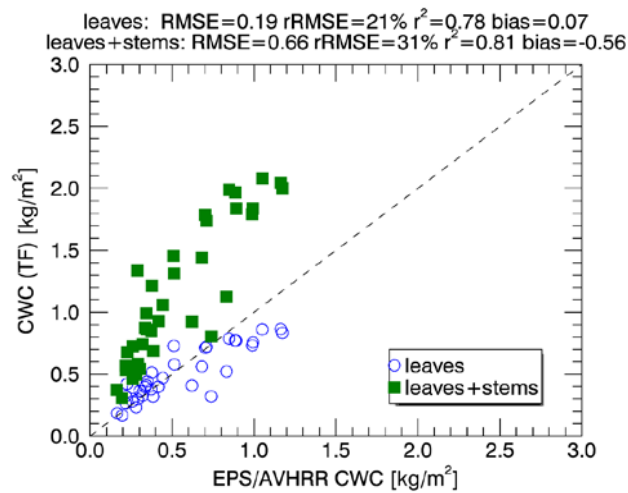

Figure 12. Scatterplot of EPS/AVHRR CWC versus aggregated TF maps for $CWC_{leaves}$ and $CWC_{leaves+stems}$, pooling the five considered ROIs for the corresponding dates of MSI acquisitions together.

Figure 13 shows the comparison between *in situ* measurements and CWC estimates derived using the LSA-SAF algorithm on MSI/Sentinel-2 data. The retrievals present a positive bias with respect to *in situ* $CWC_{leaves}$, with statistical values (considering all crop types) of RMSE=0.56 kg m$^{–2}$, $r$=0.85, and bias=0.37 kg m$^{–2}$. With regards to $CWC_{leaves+stems}$, an acceptable correspondence is observed (RMSE=0.76 kg m$^{–2}$, $r$=0.80 kg m$^{–2}$, bias= -0.30 kg m$^{–2}$), although a tendency to underestimate *in situ* measurements during high growth stages in crops with elongated stems such as corn and canola is observed.

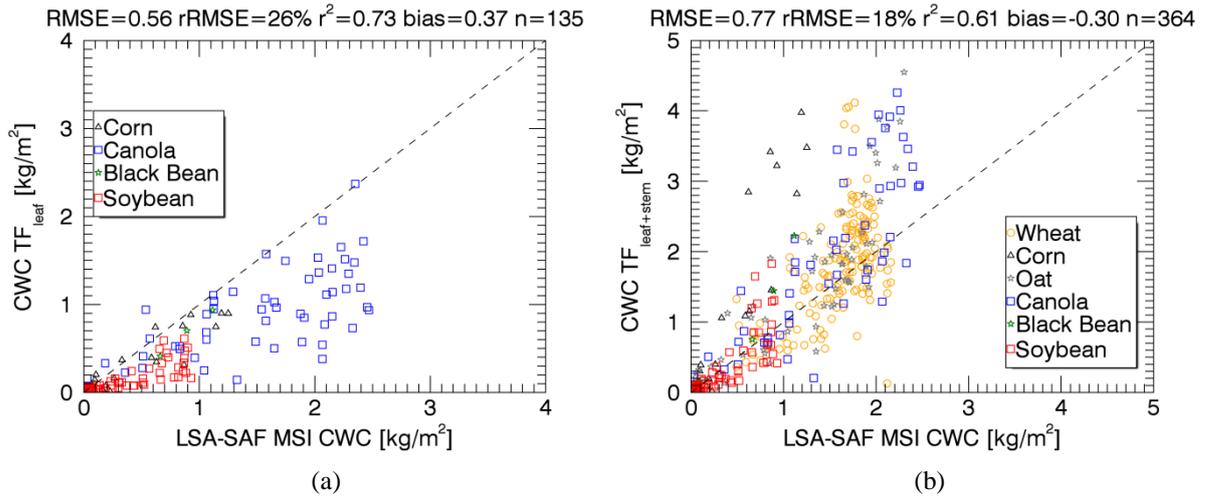

(a)                          (b)

Figure 13. Scatterplot of MSI CWC versus *in situ* measurements of CWC (CWC TF) corresponding to a) leaves, b) leaves+stems. n denotes the number of samples.

## 3.5 Accuracy assessment from time series at Dahra site

In this section, the temporal consistency of CWC derived using different instruments (MSI, MODIS and EPS/AVHRR) is assessed over a relatively homogeneous site, and compared with time series of destructive measurements (Figure 14).

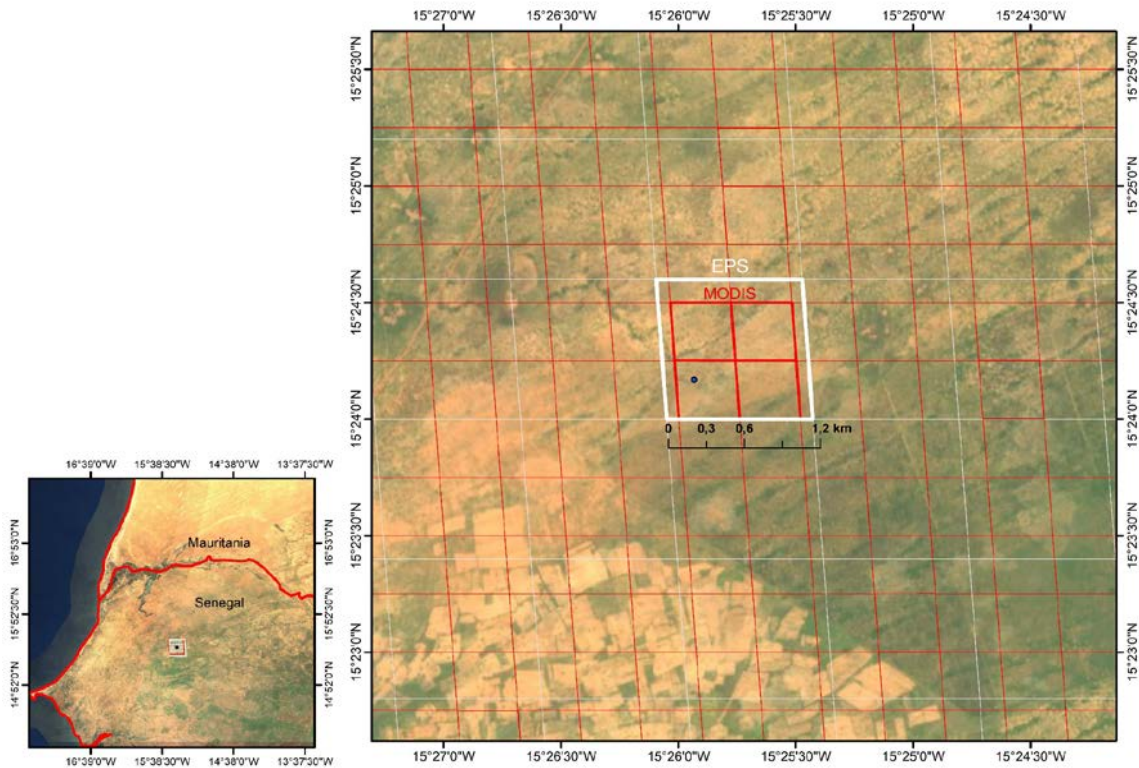

Figure 14. Left: The Dahra study area – geographical position and a 50×50 pixel EPS/AVHRR grid centred at the Dahra site (white). Right: the location of the ground station (sampled area), the MODIS-500m grid (red) and the EPS/AVHRR grid (white) (Basemap: EOX).

We observe that the LSA-SAF algorithm is capturing well the CWC phenological cycle for the Dahra field site (Figure 15(a)). No substantial differences are found when comparing MSI CWC estimates at the location of the Dahra station (pixel of 20 m resolution) with those derived using aggregated MSI observations at the EPS/AVHRR spatial resolution. The inclusion of MODIS CWC data has allowed us to extend the analysis to the 2008-2018 period too (Figure 15(b)), showing good agreement with *in situ* measurements. Still, it can be noted that MODIS profiles are hampered by missing data mainly due to cloud occurrence during the rainy season.

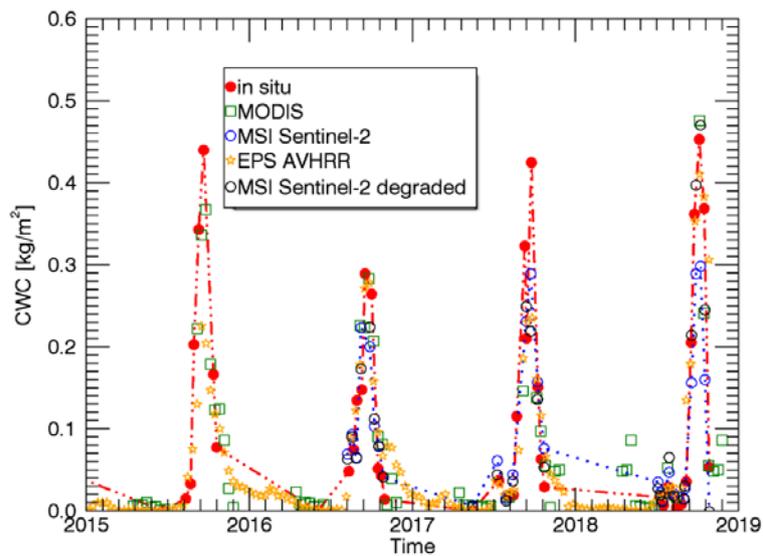

(a)

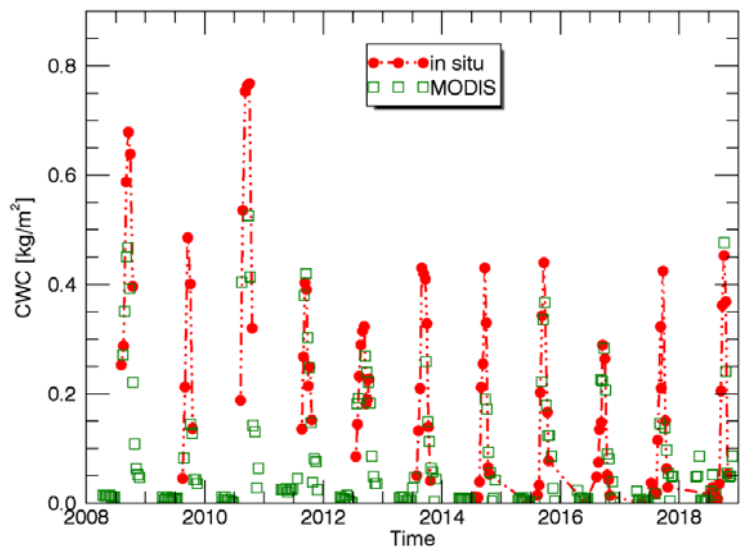

(b)

Figure 15. (a) Time series of CWC estimates from MSI (20 m), MODIS (~500 m) and EPS/AVHRR (~1.2 km) along with *in situ* measurements for the period 2015-2018. (b) Comparison over an extended period (2008-2018) using MODIS data.

The *in situ* measurements compared well with the high resolution (20 m) MSI estimates during 3 seasons (2016-2018), with RMSE=0.068 kg m$^{-2}$, $r^2$=0.84 and bias=-0.021 kg m$^{-2}$. Since the Dahra site is representative of a larger area, the assessment was achieved considering either the ~500 m resolution MODIS pixel corresponding to the biomass sampling plots and the aggregated value (~1km resolution) over a 2×2 pixel window (Figure 16). No substantial differences were found between both CWC estimates, which is indicative of the homogeneity of the site. The assessment using the aggregated (2×2 pixel window) product served us to increase the temporal sampling during the growing season (June-September) in which the MODIS product presented a high rate of missing data. The performance of the algorithm over an 11-year period is satisfactory (RMSE=0.11 kg m$^{-2}$, $r^2$=0.76, bias=-0.025 kg m$^{-2}$), particularly taking into account that CWC showed an increased variability during the extended period, with a peak at 0.7-0.8 kg m$^{-2}$ in 2010 and 2012. Despite covering a wider area, the EPS/AVHRR product during 4 years (2015-2018) was also compared with *in situ* measurements, showing statistics similar to MSI (RMSE=0.078 kg m$^{-2}$, $r^2$=0.70, bias=-0.006 kg m$^{-2}$). The above results indicate that CWC estimates from the three sensors are quite consistent and capture the general trend of *in situ* aboveground herbaceous cover. Generally, a slight overestimation of the low values and an underestimation of the high values is observed.

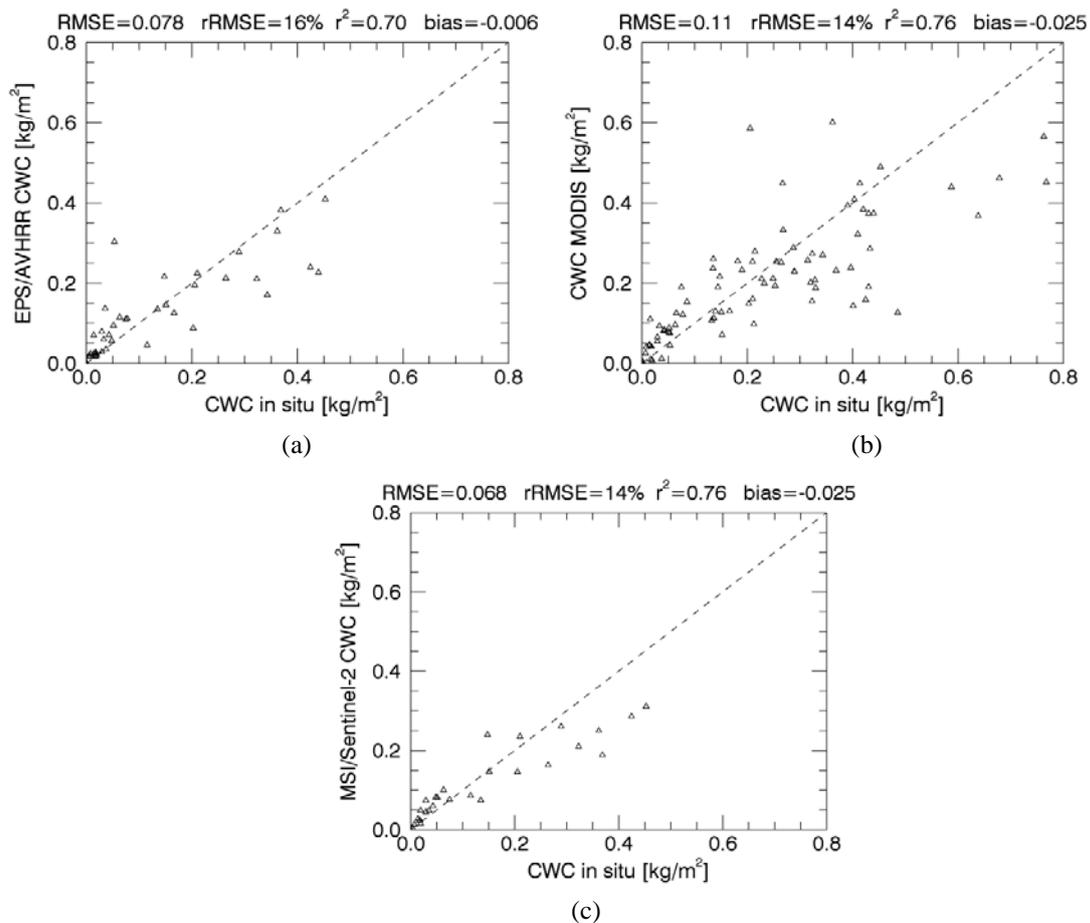

Figure 16. Scatter plot between *in situ* measurements of CWC and estimates from (a) MSI/Sentinel-2 (2016-2018), (b) MODIS (2006 and 2008-2018), (c) EPS/AVHRR (2015-2018).

## 3.6 Comparison with the MSI/SL2P algorithm

Our CWC retrievals using the LSA-SAF algorithm applied to MSI data were also directly compared with those derived from the application of the SL2P. Data points of the scatterplot (Figure 17(a)) correspond to the locations of all SMAPVEX16 measurements for the dates of the MSI overpass covering the growing cycle. We can observe a strong relationship between both estimates, which is near linear for low-to-intermediate water content levels. The scattering in the high water content range is attributable to cloudy/shadowed pixels undetected by the Quality Scene Classification layer provided by Sen2cor. Very low differences are found during the early stages of development, suggesting that both algorithms are consistent in their characterization of bare areas. However, there is a strong bias between both estimates, with the SL2P product clearly underestimating LSA-SAF observations. The comparison was also performed for the Dahra site in a multi-temporal manner using as input MSI surface reflectance (Figure 17(b)). Again, a strong bias between both approaches is observed.

A scatter plot assessing the accuracy of the SL2P algorithm against the SMAPVEX16 *in situ* data is shown in Figure S3(a) in supplementary material. In relation to TF $CWC_{leaves}$, SL2P presents a very slight negative bias (RMSE=0.30 kg m$^{-2}$, rRMSE=14%, $r^2$=0.66, bias=-0.06 kg m$^{-2}$) but it largely underestimates TF $CWC_{leaves+stems}$ (RMSE=1.25 kg m$^{-2}$, rRMSE=28%, bias= −0.95 kg m$^{-2}$). With regards to the Dahra multitemporal *in situ* measurements (Figure S3(b)), the SL2P CWC presents a substantial correlation *($r^2$=0.79)* but large differences in magnitude, with a slope of 0.15. It does not capture the seasonal variations, with a slight overestimation of the low dry season values (bias=0.04 kg m$^{-2}$) and a strong underestimation of the high rainy season values (bias=-0.35 kg m$^{-2}$).

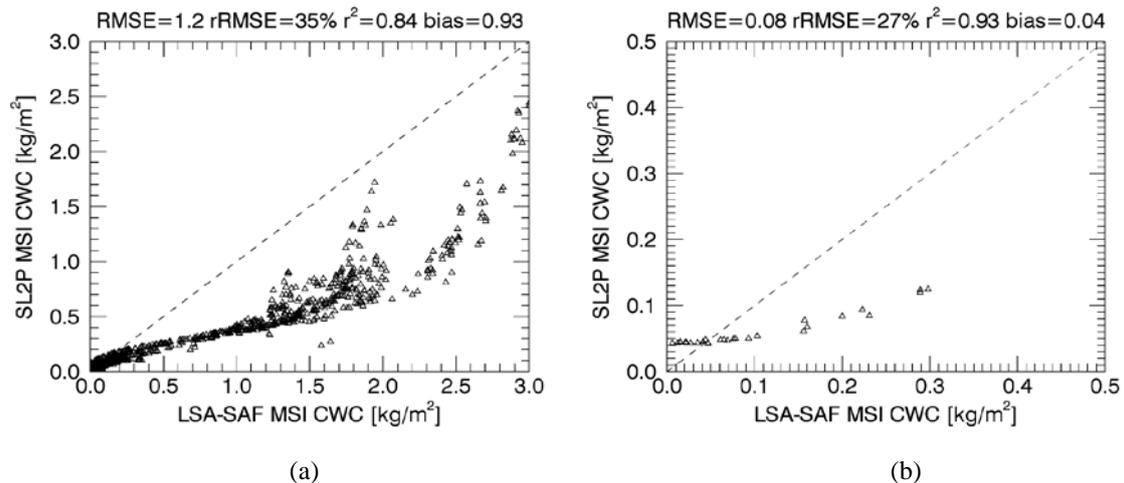

(a) (b)
Figure 17. Scatter plots of CWC estimates from MSI versus MSI/SL2P CWC product. Results correspond to a) SMAPVEX16 area; b) Dahra site.

## 3.7 Analysis of spatial scale effects

Spatial heterogeneity of vegetation within a pixel is a major issue for estimating CWC. The scale effect causes discrepancies between two estimates derived using the same algorithm but inputs at different spatial resolutions. This effect is known to influence the retrieval of variables such as LAI or CWC (unlike FVC and FAPAR), and is caused by a combination of model nonlinearity and surface heterogeneity. To study the possible impact of the spatial scale on CWC retrievals, the spatial resolution of MSI was degraded to that of EPS/AVHRR using the mean of all available high resolution (MSI) pixel values in a selected coarse resolution (EPS/AVHRR) pixel. A representative example of this comparison over the entire SMAPVEX16 area is shown in Figure 18. It can be observed that the scale effect due to a resolution change from the MSI to EPS/AVHRR does not cause any substantial bias (-0.0006 kg m$^{-2}$) and causes almost negligible differences in CWC retrievals (rRMSE=4%).

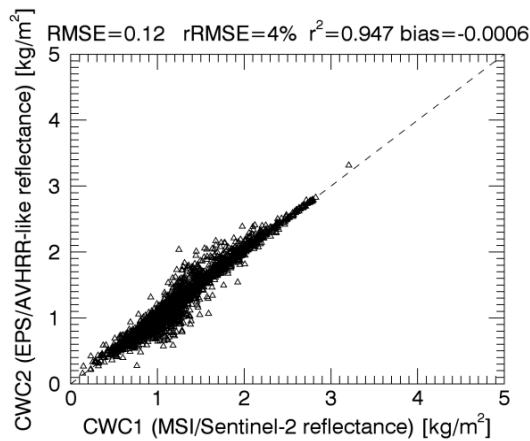

Figure 18. Scale effect on CWC for a MSI/Sentinel-2 image acquired on 30$^{th}$ September 2018 over SMAPVEX16 area (granule tiles T14UNA and T14UNV).

## 4 Discussion

One of the difficulties in retrieving CWC from optical remote sensing is that radiation hardly penetrates the canopy and CWC thereby primarily accounts for the water content of leaves in the upper layers. The EPS/AVHRR CWC product corresponds primarily to foliage elements, being rather insensitive to water from the stems and lower parts of the plant canopy. The SL2P CWC product, which is also affected by this issue, likewise accounts for water content within the foliage elements.

The number of available bands in the optical domain are 7, 8 and 13 for MODIS, Landsat and MSI/Sentinel-2, respectively. Although not all bands are equally suited to track changes in CWC, these sensors offer an enhanced level of information on canopy structure and water status. It is obvious that more available spectral bands provides more possible combinations to assess CWC from different band ratio indices, as evidenced by studies using hyperspectral data for CWC estimation (Cheng *et al*., 2006, 2008; Clevers *et al*., 2010). One limitation of the AVHRR sensor is that, even though it provides information in 6 channels, only 3 of them are useful in the context of CWC estimation. Although the retrieval of CWC from the AVHRR bands is an ill-posed problem, given the limited number of spectral channels, our findings demonstrate that the benefits of including NDII as a predictor along the reflectance are two-fold. First, it provides better constrained and more robust solutions in sparsely vegetated areas, such as shrublands during the dry season. Second, the increased sensitivity of NDII to water reduces saturation effects in dense canopies.

The LSA-SAF algorithm relies on the PROSAIL RTM which is potentially scale-dependent and moreover does not account for heterogeneity at the landscape level. However, a parameter (vCover) was introduced to account for this heterogeneity. PROSAIL considers the canopy as a collection of absorbing and scattering tissues randomly distributed in a horizontal layer (i.e., turbid medium). Casas *et al*. (2014) found that the use of PROSAIL for CWC estimation over forest, shrubs, and grassland was limited without precise ground information. In order to alleviate this problem, the LSA-SAF algorithm extracts *a priori* knowledge from a large database of in situ measured leaf traits, which is used in the PROSAIL simulations.

The influence of spatial resolution on LSA-SAF algorithm has shown to be minimal, with very small differences in CWC retrievals (rRMSE=4%) and negligible bias. Furthermore, EPS/AVHRR CWC values are generally consistent with those obtained by applying the LSA-SAF algorithm on data of higher spatial resolution such as MSI/Sentinel-2 and MODIS/Terra & Aqua. CWC estimates from MODIS and MSI were also highly consistent, demonstrating that the algorithm is sensor agnostic.

The LSA-SAF uncertainty requirements for EPS/AVHRR CWC products are Maximum [0.3 kg m$^{-2}$, 35%] for target accuracy and Maximum [0.25 kg m$^{-2}$, 30%] for optimal accuracy. The present study performs a validation of EPS/AVHRR products in two study areas. The observations show a close agreement with *in situ* CWC$_{leaves}$ measurements in SMAPVEX16 (RMSE = 0.19 kg m$^{-2}$), demonstrating only a slight overestimation for high water content values (mean bias error = 0.07). Similarly, in the Dahra site the EPS/AVHRR product captures the general trend of *in situ* measurements of CWC for all aboveground herbaceous cover, although with a slight underestimation for high values. Optimal accuracy requirements are clearly achieved (e.g. RMSE = 0.068 kg m$^{-2}$) even if the vegetation variability is very limited.

The comparison of the application of our LSA-SAF algorithm with the original SL2P algorithm on the same MSI/Sentinel-2 dataset shows a strong relationship, which is almost linear for low and intermediate CWC levels. Large differences are however found between both estimates at high CWC levels, which is likely due to differences in the characterization of the leaf/canopy attributes in PROSAIL and in the inputs used: 3 bands plus a water index (LSA-SAF) versus 8 MSI/Sentinel-2 bands plus the cosine of the sun zenith angle, view zenith angle, and relative azimuth angle (MSI/SL2P).

Over SMAPVEX16, results from comparing the LSA-SAF algorithm for both MSI/Sentinel-2 and MODIS at various resolutions (original and EPS/AVHRR-like) with ground reference data are, in general, highly consistent. Some discrepancies exist, partly due to differences in their compositing period and spatial resolution, which limit the direct comparability between retrievals obtained from the different sensor. In particular, LSA-SAF MSI CWC retrievals underestimate the ground-truth $CWC_{leaves+stems}$ values observed for corn and canola. This is not surprising because stems for these crops at the end of the growing season are considerably larger than leaves and sensitivity to their contribution is therefore reduced. Still, there is a tendency to overestimate foliage water content ($CWC_{leaves}$), whereas a reasonable correspondence with $CWC_{leaves+stems}$ in certain crops is observed. This result is not consistent with the higher sensitivity of the EPS/AVHRR product to $CWC_{leaves}$ and could be due to differences in sensors Point Spread Functions (PSFs), processing chains (atmospheric correction, temporal compositing and BRDF normalization), spatial resolution and coregistration errors. The results suggest that the coarser resolution of the EPS/AVHRR reflectance presents a poorer quality in relation to both MSI and MODIS, which may reduce the sensitivity to the water signal from the stems.

The MSI/SL2P product over SMAPVEX16 shows a closer agreement with $CWC_{leaves}$ while it strongly underestimates $CWC_{leaves+stems}$. These results are in line with an underestimation of SL2P CWC reported by Djamaia *et al.*, (2019) over the same site considered here. A poorer performance of the MSI/SL2P algorithm is found for the Dahra site, showing a slight positive offset at the early stages of the growing cycle and unrealistic low variations during the entire cycle. Since machine learning methods are known to provide unrealistic values if observations deviate from the simulations, this result suggests that the SL2P algorithm may fail to properly characterize the soil and leaf/canopy attributes in this semi-arid region.

The LSA-SAF algorithm provides global EPS/AVHRR estimates for the following six variables: LAI, FVC, FAPAR, CWC, $C_w$ and $C_{dm}$. Although $C_w$ and $C_{dm}$ are not products to be disseminated to users, they may have applicability for the assessment of the status of vegetation stress conditions and have served us to indirectly assess the validity of the algorithm. For example, the scatterplot in Figure S4 (see supplementary material) combines three variables joinlty retrieved by the hybrid algorithm, i.e. CWC, LAI and $C_w$. The product $C_w \cdot LAI$ shows a close agreement with CWC (rRMSE=1.2%), which is coherent with the CWC definition, revaling that the retrieval algorithm is well constrained.

# 5   Conclusions

We present the LSA-SAF algorithm to operationally produce estimates of CWC at a global scale based on optical data from EPS/AVHRR satellites. It uses an accurate representation of the PROSAIL parameter distribution and introduces the NDII water index as a predictor for the joint retrieval of CWC and other EPS/AVHRR variables (LAI, FAPAR, FVC) using a single machine learning algorithm based on multi-output Gaussian processes. The 1.1 km 10-day EPS/AVHRR CWC product is generated at a global scale, providing pixel-wise uncertainty estimates and quality flag information to identify unreliable observations. Because of the poor sensitivity of the AVHRR signal to water from the lower parts of the plant, we show EPS AVHRR CWC product sensitivity corresponds primarily to foliage elements.

Despite the limited number of input bands of the AVHRR sensor, the algorithm has proven to be effective to estimate CWC and its temporal variations. The inclusion of the NDII water index as predictor is shown to be important to avoid the presence of anomalous temporal patterns during the dry season in sparse areas and mitigate saturation effects in dense canopies. The following conclusions can be drawn about the feasibility of the LSA-SAF algorithm:

(i) The CWC retrievals are virtually invariant when degrading the data from high to coarse spatial resolutions, thus no bias would be expected from its application to coarse resolution satellite data such as EPS/AVHRR.

(ii) The LSA-SAF algorithm is sensor agnostic showing consistent CWC retrievals when applied on reflectances measured by MSI/Sentinel-2 (20 m) and MODIS (500 m).

(iii) The intercomparison between the LSA-SAF and the SL2P algorithms using identical MSI/Sentinel-2 data sets shows a strong (near linear) relationship with however a large bias towards higher CWC estimates.

The main limitation for the validation of the EPS/AVHRR CWC products is the lack of ground data representative of coarse spatial resolution and the absence of similar global CWC satellite products. The results of the present study suggest that the LSA-SAF EPS/AVHRR algorithm is robust, as confirmed through numerical comparisons with multitemporal ground measurements from SMAPVEX16 in Canada (croplands) and Dahra (semi-arid savanna grassland) in Senegal, with RMSE values of 0.21 kg m$^{-2}$ and 0.09 kg m$^{-2}$, respectively. However, the limited data sets used in this assessment are not representative of all terrestrial biomes but only some croplands and, therefore, further validation research is required in future work, mainly over forests where the PROSAIL turbid assumption does not hold properly (Casas *et al*., 2014; Verger *et al*., 2011).

The flexibility of the methodology enables products to be derived using data from higher spatial and spectral resolution missions such as the future EPS-SG/VII and EPS-SG/3MI EUMETSAT sensors.

This will lead to enhanced products of CWC, which may lead to improvements in vegetation monitoring and have implications in ecosystem-climate interaction studies and modeling.

## ACKNOWLEDGMENTS

Funding support by LSA-SAF (EUMETSAT) and ESCENARIOS (CGL2016-75239-R) projects is acknowledged. Tagesson was funded by the Swedish National Space Board (SNSB Dnr 95/16). This research was also financially supported by the NASA Earth Observing System MODIS project (grant NNX08AG87A) and by the European Research Council (ERC) funding under the ERC Consolidator Grant 2014 SEDAL project under Grant Agreement 647423 and LEAVES project RTI2018-096765-A-100 (MCIU/AEI/FEDER, UE). We also thank the anonymous reviewers for their valuable comments, which enhanced the quality of the paper.

# References

Asner, G. P., Brodrick, P. G., Anderson, C. B., Vaughn, N., Knapp, D. E., & Martin, R. E. (2016). Progressive forest canopy water loss during the 2012–2015 California drought. *Proceedings of the National Academy of Sciences, 113*(2), E249-E255.

Baret, F., Hagolle, O., Geiger, B., Bicheron, P., Miras, B., Huc, M., *et al*. (2007). LAI, fAPAR and fCover CYCLOPES global products derived from VEGETATION: Part 1: Principles of the algorithm. *Remote sensing of environment, 110*(3), 275–286.

Baret, F., & Buis, S. (2008). Estimating canopy characteristics from remote sensing observations: Review of methods and associated problems. In *Advances in land remote Sensing* (pp. 173-201). Springer, Dordrecht.

Berger, K., Atzberger, C., Danner, M., D'Urso, G., Mauser, W., Vuolo, F., & Hank, T. (2018). Evaluation of the PROSAIL model capabilities for future hyperspectral model environments: a review study. *Remote Sensing, 10*(1), 85.

Breiman, L. (2001). Random forests. Machine learning, 45(1), 5–32.

Camacho, Baret, F., Weiss, M., Fernandes, R., Berthelot, B., Sánchez, J., Latorre, C., García-Haro, F.J., Duca, R., (2013), Validación de algoritmos para la obtenció de variables biofísicas con datos Sentinel2 en la ESA: Proyecto VALSE-2. XV *Congreso de la Asociación Española de Teledetección*,INTA, Torrejón de Ardoz (Madrid). 22-24 Oct., 2013.

Camacho, F., Lacaze, R., Latorre, C., Baret, F., De la Cruz, F., Demarez, V., *et al*. (2014). A network of sites for ground biophysical measurements in support of Copernicus Global Land Product Validation. In *Proceedings of the IV RAQRS conference*, Torrent (pp. 22-26).

Campos-Taberner, M., García-Haro, F. J., Camps-Valls, G., Grau-Muedra, G., Nutini, F., Crema, A., & Boschetti, M. (2016). Multitemporal and multiresolution leaf area index retrieval for operational local rice crop monitoring. *Remote Sensing of Environment, 187*, 102-118.

Campos-Taberner, M., Moreno-Martínez, Á., García-Haro, F., Camps-Valls, G., Robinson, N., Kattge, J., & Running, S. (2018). Global estimation of biophysical variables from google earth engine platform. *Remote Sensing, 10*(8), 1167.

Camps-Valls, G., Sejdinovic, D., Runge, J., & Reichstein, M. (2019). A perspective on Gaussian processes for Earth observation. *National Science Review*. In press. Volume 6, Issue 4, July 2019, Pages 616–618, https://doi.org/10.1093/nsr/nwz028.

Camps-Valls, G., Verrelst, J., Munoz-Mari, J., Laparra, V., Mateo-Jimenez, F., & Gomez-Dans, J. (2016). A survey on gaussian processes for earth-observation data analysis: A comprehensive investigation. *IEEE Geoscience and Remote Sensing Magazine, 4*(2), 58–78.

Carter, G.A. Responses of leaf spectral reflectance to plant stress. (1993) *American Journal of Botany, 80(3),* 239–243.

Casas, A., Riaño, D., Ustin, S. L., Dennison, P., & Salas, J. (2014). Estimation of water-related biochemical and biophysical vegetation properties using multitemporal airborne hyperspectral data and its comparison to MODIS spectral response. *Remote Sensing of Environment, 148*, 28-41.

Ceccato, P., Flasse, S., Tarantola, S., Jacquemoud, S., & Grégoire, J. M. (2001). Detecting vegetation leaf water content using reflectance in the optical domain. *Remote sensing of environment, 77*(1), 22–33.

Ceccato, P., Flasse, S., & Gregoire, J. M. (2002). Designing a spectral index to estimate vegetation water content from remote sensing data: Part 2. Validation and applications. *Remote Sensing of Environment, 82*(2-3), 198–207.

Cheng, Y. B., Zarco-Tejada, P. J., Riaño, D., Rueda, C. A., & Ustin, S. L. (2006). Estimating vegetation water content with hyperspectral data for different canopy scenarios: Relationships between AVIRIS and MODIS indexes. *Remote Sensing of Environment, 105*(4), 354–366.

Cheng, Y. B., Ustin, S. L., Riaño, D., & Vanderbilt, V. C. (2008). Water content estimation from hyperspectral images and MODIS indexes in Southeastern Arizona. *Remote Sensing of Environment, 112*(2), 363–374.

Chowdhury, E. H., & Hassan, Q. K. (2015). Operational perspective of remote sensing-based forest fire danger forecasting systems. *ISPRS Journal of Photogrammetry and Remote Sensing, 104*, 224–236.

Chuvieco, E., Wagtendonk, J., Riaño, D., Yebra, M., & Ustin, S. L. (2009). Estimation of fuel conditions for fire danger assessment. In Earth observation of wildland fires in Mediterranean ecosystems (pp. 83-96). Springer, Berlin, Heidelberg.

Clevers, J. G., Kooistra, L., & Schaepman, M. E. (2010). Estimating canopy water content using hyperspectral remote sensing data. *International Journal of Applied Earth Observation and Geoinformation, 12*(2), 119–125.

Combal, B., Baret, F., Weiss, M., Trubuil, A., Mace, D., Pragnere, A. *et al.* (2003). Retrieval of canopy biophysical variables from bidirectional reflectance: Using prior information to solve the ill-posed inverse problem. *Remote sensing of environment, 84*(1), 1-15.

Cosh, M. H., White, W. A., Colliander, A., Jackson, T. J., Prueger, J. H., Hornbuckle, B. K.,*et al.* (2019). Estimating vegetation water content during the Soil Moisture Active Passive Validation Experiment 2016. *Journal of Applied Remote Sensing, 13*(1), 014516.

Djamai, N., Fernandes, R., Weiss, M., McNairn, H., & Goïta, K. (2019). Validation of the Sentinel Simplified Level 2 Product Prototype Processor (SL2P) for mapping cropland biophysical variables using Sentinel-2/MSI and Landsat-8/OLI data. *Remote sensing of environment, 225*, 416–430.

Drusch, M., Del Bello, U., Carlier, S., Colin, O., Fernandez, V., Gascon, F., *et al*. (2012). Sentinel-2: ESA's optical high-resolution mission for GMES operational services. *Remote sensing of Environment, 120*, 25-36.

Entekhabi, D., Njoku, E. G., O'Neill, P. E., Kellogg, K. H., Crow, W. T., Edelstein, W. N., *et al*. (2010). The Soil Moisture Active Passive (SMAP) mission. Proceedings of the IEEE, 98(5), 704-716. doi: 10.1109/JPROC.2010.2043918.

Fensholt, R., Sandholt, I., & Rasmussen, M. S. (2004). Evaluation of MODIS LAI, fAPAR and the relation between fAPAR and NDVI in a semi-arid environment using in situ measurements. *Remote sensing of Environment, 91*(3-4), 490–507.

Fernandes, R., Plummer, S., Nightingale, J., Baret, F., Camacho, F., Fang, H., Garrigues, S., Gobron, N., Lang, M., Lacaze, R., LeBlanc, S., Meroni, M., Martinez, B., Nilson, T., Pinty, B., Pisek, J., Sonnentag, O., Verger, A., Welles, J., Weiss, M., & Widlowski, J.L. (2014). Global Leaf Area Index Product Validation Good Practices. Version 2.0. In G. Schaepman-Strub, M. Román, & J. Nickeson (Eds.),*Best Practice for Satellite-Derived Land Product Validation* (p. 76): Land Product Validation Subgroup (WGCV/CEOS), doi:10.5067/doc/ceoswgcv/lpv/lai.002

Friedl, M. A., Sulla-Menashe, D., Tan, B., Schneider, A., Ramankutty, N., Sibley, A., & Huang, X. (2010). MODIS Collection 5 global land cover: Algorithm refinements and characterization of new datasets. *Remote Sensing of Environment, 114*(1), 168–182.

García-Haro, F. J., Campos-Taberner, M., Muñoz-Marí, J., Laparra, V., Camacho, F., Sánchez-Zapero, J., & Camps-Valls, G. (2018). Derivation of global vegetation biophysical parameters from EUMETSAT Polar System. *ISPRS journal of photogrammetry and remote sensing, 139*,

57–74.

Geiger, B., Carrer, D., Hautecoeur, O., Franchistéguy, L., Roujean, J.-L., Meurey, C., Ceamanos, X., Jacob, G. (2018). Algorithm Theoretical Basis Document (ATBD) for Ten-day Surface Albedo from EPS/Metop/AVHRR (ETAL). Retrieved from https://landsaf.ipma.pt/GetDocument.do?id=641.

Hantson, S., Arneth, A., Harrison, S. P., Kelley, D. I., Prentice, I. C., Rabin, S. S. *et al.* (2016). The status and challenge of global fire modelling. *Biogeosciences*, *13*(11), 3359–3375.

Hardisky, M. A., Klemas, V., & Smart, M. (1983). The influence of soil salinity, growth form, and leaf moisture on the spectral radiance of. *Spartina alterniflora, 49,* 77–83.

Irmak, S., Haman, D. Z., & Bastug, R. (2000). Determination of crop water stress index for irrigation timing and yield estimation of corn. *Agronomy Journal*, *92*(6), 1221–1227.

Jacquemoud, S., & Baret, F. (1990). PROSPECT: A model of leaf optical properties spectra. *Remote sensing of environment, 34*(2), 75–91.

Jackson, T.J., & Schmugge T.J. (1991). Vegetation effects on the microwave emission of soils, *Remote Sensing of Environment, 36(3)*, 203-212,

Jackson, T. J., Chen, D., Cosh, M., Li, F., Anderson, M., Walthall, C., *et al.* (2004). Vegetation water content mapping using Landsat data derived normalized difference water index for corn and soybeans. *Remote Sensing of Environment, 92*(4), 475–482.

Johnson, J. E., Laparra, V., & Camps-Valls, G. (2019). Accounting for Input Noise in Gaussian Process Parameter Retrieval. *IEEE Geoscience and Remote Sensing Letters*. In press

Kattge, J., Diaz, S., Lavorel, S., Prentice, I. C., Leadley, P., Bönisch, G., *et al.* (2011). TRY–a global database of plant traits. *Global change biology, 17*(9), 2905–2935.

Konings, A.G., M. Piles, N. Das, D. Entekhabi (2017). L-Band Vegetation Optical Depth and Effective Scattering Albedo Estimation from SMAP, *Remote Sensing of the Environment, 198*, 460–470.

Liu, Y. Y., Van Dijk, A. I., De Jeu, R. A., Canadell, J. G., McCabe, M. F., Evans, J. P., and Wang, G. (2015): Recent reversal in loss of global terrestrial biomass, *Nature Climate Change, 5*, 470–474.

Martin, R.E., Asner, G. P., Francis, E., Ambrose, A., Baxter, W., Das, A. J., *et al.* (2018). Remote measurement of canopy water content in giant sequoias (Sequoiadendron giganteum) during drought. *Forest Ecology and Management, 419*, 279–290.

Martínez, B., Sanchez-Ruiz, S., Gilabert, M. A., Moreno, A., Campos-Taberner, M., García-Haro, F. J., *et al.* (2018). Retrieval of daily gross primary production over Europe and Africa from an ensemble of SEVIRI/MSG products. *International journal of applied earth observation and geoinformation, 65,* 124–136.

Martínez, B., Gilabert, M.A., Sánchez-Ruiz, S., Campos-Taberner, M., García-Haro, F.J., Brüemmer, C., Carrara, A., Feig, G., Grünwald, T., Mammarella, I., Tagesson, T. (2019). Evaluation of the LSA-SAF Gross Primary Production product derived from SEVIRI/MSG data (MGPP). *ISPRS Journal of Photogrammetry and Remote Sensing*, In review.

Mbow, C., Fensholt, R., Rasmussen, K., & Diop, D. (2013). Can vegetation productivity be derived from greenness in a semi-arid environment? Evidence from ground-based measurements. *Journal of Arid Environments, 97*, 56–65.

McNairn, H., K. Gottfried, and J. Powers. 2018. SMAPVEX16 Manitoba CropScan Data, Version 1. Vegetation Water Content. Boulder, Colorado USA. NASA National Snow and Ice Data Center Distributed Active Archive Center. https://doi.org/10.5067/Y4W64RE5RWBF.

Moghaddam, M., & Saatchi, S. S. (1999). Monitoring tree moisture using an estimation algorithm applied to SAR data from BOREAS. *IEEE Transactions on Geoscience and Remote Sensing, 37*(2), 901–916.

Morisette, J. T., Baret, F., Privette, J. L., Myneni, R. B., Nickeson, J. E., Garrigues, S., Shabanov, N. V., Weiss, M., Fernandes, R. A., Leblanc, S. G., Kalacska, M., Sanchez-Azofeifa, G. A., Chubey, M., Rivard, B., Stenberg, P., Rautiainen, M., Voipio, P., Manninen, T., Pilant, A. N., Lewis, T. E., Iiames, J. S., Colombo, R., Meroni, M., Busetto, L., Cohen, W. B., Turner, D. P., Warner, E. D., Petersen, G. W., Seufert, G. and Cook, R. (2006). Validation of global moderate-resolution LAI products: A framework proposed within the CEOS Land Product Validation subgroup. *IEEE Transactions on Geoscience and Remote Sensing*. 44(7): 1804-1817.

Nilsen, E. T., & Orcutt, D. M. (1996). Physiology of plants under stress. Abiotic factors. *Physiology of plants under stress. Abiotic factors*.

Parzen, E. (1962). On estimation of a probability density function and mode. *The annals of mathematical statistics, 33*(3), 1065–1076.

Pasqualotto, N., Delegido, J., Van Wittenberghe, S., Verrelst, J., Rivera, J. P., & Moreno, J. (2018). Retrieval of canopy water content of different crop types with two new hyperspectral indices: Water Absorption Area Index and Depth Water Index. *International journal of applied earth observation and geoinformation, 67*, 69–78.

Peñuelas, J., Pinol, J., Ogaya, R., & Filella, I. (1997). Estimation of plant water concentration by the reflectance water index WI (R900/R970). I*nternational Journal of Remote Sensing, 18*(13), 2869–2875.

Rahimzadeh-Bajgiran, P., Omasa, K., & Shimizu, Y. (2012). Comparative evaluation of the Vegetation Dryness Index (VDI), the Temperature Vegetation Dryness Index (TVDI) and the improved TVDI (iTVDI) for water stress detection in semi-arid regions of Iran. *ISPRS Journal of Photogrammetry and Remote Sensing, 68*, 1–12.

Rasmussen, M. O., Göttsche, F. M., Diop, D., Mbow, C., Olesen, F. S., Fensholt, R., & Sandholt, I. (2011). Tree survey and allometric models for tiger bush in northern Senegal and comparison with tree parameters derived from high resolution satellite data. *International Journal of Applied Earth Observation and Geoinformation, 13*(4), 517–527.

Riaño, D., Vaughan, P., Chuvieco, E., Zarco-Tejada, P. J., & Ustin, S. L. (2005). Estimation of fuel moisture content by inversion of radiative transfer models to simulate equivalent water thickness and dry matter content: Analysis at leaf and canopy level. *IEEE Transactions on Geoscience and Remote Sensing, 43*(4), 819–826.

Rossini, M., Fava, F., Cogliati, S., Meroni, M., Marchesi, A., Panigada, C., *et al*. (2013). Assessing canopy PRI from airborne imagery to map water stress in maize. *ISPRS Journal of Photogrammetry and Remote Sensing, 86*, 168–177.

Roujean, J. L., Leroy, M., Deschamps, P. Y. (1992). A bidirectional reflectance model of the Earth's surface for the correction of remote sensing data. *Journal of Geophysical Research: Atmospheres, 97*(D18), 20455–20468.

Tagesson, T., Fensholt, R., Guiro, I., Rasmussen, M. O., Huber, S., Mbow, C., *et al*. (2015). Ecosystem properties of semiarid savanna grassland in West Africa and its relationship with environmental variability. *Global change biology, 21*(1), 250–264.

Trigo, I. F., Dacamara, C. C., Viterbo, P., Roujean, J. L., Olesen, F., Barroso, C., *et al*. (2011). The satellite application facility for land surface analysis. *International Journal of Remote Sensing, 32*(10), 2725–2744.

Trombetti, M., Riaño, D., Rubio, M. A., Cheng, Y. B., & Ustin, S. L. (2008). Multi-temporal vegetation canopy water content retrieval and interpretation using artificial neural networks for the continental USA. *Remote Sensing of Environment, 112*(1), 203–215.

Ullah, S., Skidmore, A. K., Ramoelo, A., Groen, T. A., Naeem, M., & Ali, A. (2014). Retrieval of leaf water content spanning the visible to thermal infrared spectra. *ISPRS journal of photogrammetry and remote sensing, 93,* 56–64.

Verger, A., Baret, F., Camacho, F. (2011). Optimal modalities for radiative transfer-neural network estimation of canopy biophysical characteristics: Evaluation over an agricultural area with CHRIS/PROBA observations. *Remote Sensing of Environment, 115*(2), 415–426.

Verhoef, W. (1984). Light scattering by leaf layers with application to canopy reflectance modeling: The SAIL model. *Remote Sensing of Environment*, *16*(2), 125–1141.

Weiss, M., Baret, F., Myneni, R., Pragnère, A., & Knyazikhin, Y. (2000). Investigation of a model inversion technique to estimate canopy biophysical variables from spectral and directional reflectance data. *Agronomie, 20*(1), 3–22.

Weiss, M., & Baret, F. (2016). S2 ToolBox level 2 products. Version 1.1. Available on line at http://step.esa.int/docs/extra/ATBD_S2ToolBox_L2B_V1.1.pdf.

Xiao, X., Boles, S., Liu, J., Zhuang, D., Frolking, S., Li, C., *et al.* (2005). Mapping paddy rice agriculture in southern China using multi-temporal MODIS images. *Remote sensing of environment, 95*(4), 480–492.

Yebra, M., Dennison, P. E., Chuvieco, E., Riano, D., Zylstra, P., Hunt Jr, E. R., *et al.* (2013). A global review of remote sensing of live fuel moisture content for fire danger assessment: Moving towards operational products. *Remote Sensing of Environment, 136*, 455–468.

Yebra, M., & Chuvieco, E. (2009). Generation of a species-specific look-up table for fuel moisture content assessment. *IEEE Journal of Selected Topics in Applied Earth Observations and Remote Sensing, 2*(1), 21–26.

Yebra, M., Quan, X., Riaño, D., Larraondo, P. R., van Dijk, A. I., & Cary, G. J. (2018). A fuel moisture content and flammability monitoring methodology for continental Australia based on optical remote sensing. *Remote Sensing of Environment, 212*, 260–272.

Zarco-Tejada, P. J., Rueda, C. A., & Ustin, S. L. (2003). Water content estimation in vegetation with MODIS reflectance data and model inversion methods. *Remote sensing of Environment, 85*(1), 109–124.

Žežula, I. (2009). On multivariate Gaussian copulas. *Journal of Statistical Planning and Inference, 139*(11), 3942–3946.

Zhang, F. and G Zhou, (2015), Estimation of Canopy Water Content by Means of Hyperspectral Indices Based on Drought Stress Gradient Experiments of Maize in the North Plain China, *Remote Sensing, 7*(11), 15203–15223.

Zhu, X., Wang, T., Darvishzadeh, R., Skidmore, A. K., & Niemann, K. O. (2015). 3D leaf water content mapping using terrestrial laser scanner backscatter intensity with radiometric correction. *ISPRS journal of photogrammetry and remote sensing, 110,* 14–23.